\documentclass[preprint2]{aastex}
\usepackage{amsmath} 
\usepackage{amssymb} 
\usepackage{amsthm}
\usepackage{hyperref}

\slugcomment{Accepted in Astrophysical Journal}
\shorttitle{The Submm Pol-spectrum of M17}
\shortauthors{Zeng et al.}

\begin{document}

%% LaTeX will automatically break titles if they run longer than
%% one line. However, you may use \\ to force a line break if
%% you desire.

\title{The Submillimeter Polarization Spectrum of M17}

%% Use \author, \affil, and the \and command to format
%% author and affiliation information.
%% Note that \email has replaced the old \authoremail command
%% from AASTeX v4.0. You can use \email to mark an email address
%% anywhere in the paper, not just in the front matter.
%% As in the title, use \\ to force line breaks.

\author{Lingzhen Zeng\altaffilmark{1,2}, Charles L. Bennett\altaffilmark{2}, Nicholas L. Chapman\altaffilmark{3}, David T. Chuss\altaffilmark{4}, Izaskun Jimenez-Serra\altaffilmark{1}, Giles Novak\altaffilmark{3} and John E. Vaillancourt\altaffilmark{5}}

\altaffiltext{1}{Harvard-Smithsonian Center for Astrophysics, 60 Garden Street, Cambridge, MA 02138; \email{lingzhen@cfa.harvard.edu}}
\altaffiltext{2}{Department of Physics and Astronomy, Johns Hopkins University, 3400 North Charles Street, Baltimore, MD 21218}
\altaffiltext{3}{Center for Interdisciplinary Exploration and Research in Astrophysics (CIERA) \& Department of Physics and Astronomy, Northwestern University, 2145 Sheridan Road, Evanston, IL 60208}
\altaffiltext{4}{Observational Cosmology Laboratory, NASA Goddard Space Flight Center, Code 665, Greenbelt, MD 20771}
\altaffiltext{5}{SOFIA Science Center, Universities Space Research Association, NASA Ames Research Center, MS 232-11, Moffett Field, CA 94035}

%% Mark off your abstract in the ``abstract\arcsec environment. In the manuscript
%% style, abstract will output a Received/Accepted line after the
%% title and affiliation information. No date will appear since the author
%% does not have this information. The dates will be filled in by the
%% editorial office after submission.

\begin{abstract}
We present 450 $\mu$m polarimetric observations of the M17 molecular cloud obtained with the SHARP polarimeter at the Caltech Submillimeter Observatory. Across the observed region, the magnetic field orientation is consistent with previous submillimeter and far-infrared polarization measurements. Our observations are centered on a region of the molecular cloud that has been compressed by stellar winds from a cluster of OB stars. We have compared these new data with previous 350 $\mu$m polarimetry and find an anti-correlation between the 450 to 350 $\mu$m polarization magnitude ratio and the ratio of 21 cm to 450 $\mu$m intensity. The polarization ratio is lower near the east end of the studied region where the cloud is exposed to stellar winds and radiation. At the west end of the region, the polarization ratio is higher. We interpret the varying polarization spectrum as evidence supporting the radiative alignment torque (RAT) model for grain alignment, implying higher alignment efficiency in the region that is exposed to a higher anisotropic radiation field.
\end{abstract}

%% Keywords should appear after the \end{abstract} command. The uncommented
%% example has been keyed in ApJ style. See the instructions to authors
%% for the journal to which you are submitting your paper to determine
%% what keyword punctuation is appropriate.


%% From the front matter, we move on to the body of the paper.
%% In the first two sections, notice the use of the natbib \citep
%% and \citet commands to identify citations.  The citations are
%% tied to the reference list via symbolic KEYs. The KEY corresponds
%% to the KEY in the \bibitem in the reference list below. We have
%% chosen the first three characters of the first author\arcmins name plus
%% the last two numeral of the year of publication as our KEY for
%% each reference.

%% Authors who wish to have the most important objects in their paper
%% linked in the electronic edition to a data center may do so by tagging
%% their objects with \objectname{} or \object{}.  Each macro takes the
%% object name as its required argument. The optional, square-bracket 
%% argument should be used in cases where the data center identification
%% differs from what is to be printed in the paper.  The text appearing 
%% in curly braces is what will appear in print in the published paper. 
%% If the object name is recognized by the data centers, it will be linked
%% in the electronic edition to the object data available at the data centers  
%%
%% Note that for sources with brackets in their names, e.g. [WEG2004] 14h-090,
%% the brackets must be escaped with backslashes when used in the first
%% square-bracket argument, for instance, \object[\[WEG2004\] 14h-090]{90}).
%%  Otherwise, LaTeX will issue an error. 

\section{Introduction}
\label{sec:intrm17}

Magnetic fields are believed to play an important role in the dynamics and evolution of galactic molecular clouds and hence affect the star forming processes therein. Since the dust temperature of a typical molecular cloud is $\sim$ 10 -- 20 K, the submillimeter part of the electromagnetic spectrum is a very important window for studying the physics of star formation. Submillimeter polarimetry provides one of the best tools for mapping interstellar magnetic fields in star forming regions \citep{1999ApJ...520..706C}, because asymmetric dust grains are partially aligned by magnetic fields. The physics of the alignment process is an active area of research.

There are several theoretical models for magnetic alignment of interstellar dust grains (see reviews by \citealp{2003JQSRT..79..881L, 2007JQSRT.106..225L}). Among them, the ``radiative alignment torques'' (RAT) model is the most favored candidate. In this model, photons from an anisotropic radiation field produce a net radiative alignment torque on irregularly shaped grains, because the grains present different cross sections to right- and left-handed circularly polarized photons \citep{1972Ap&SS..18..337D, 1976Ap&SS..43..257D, 1996ApJ...470..551D, 1997ApJ...480..633D, 2003JQSRT..79..881L, 2007JQSRT.106..225L}. As in the case for other grain alignment theories, the grain axis with the largest moment of inertia is aligned parallel to the spin axis, and furthermore the spin axis is aligned with the local magnetic field. Since the grains will emit and absorb most efficiently along the long grain axis, polarization is observed perpendicular to the magnetic field in emission, but parallel to the field in absorption (or extinction). As a result, measurement of the direction of polarization provides knowledge of the orientation of the interstellar magnetic field, as projected onto the plane of the sky \citep{2003JQSRT..79..881L, 2007JQSRT.106..225L}.

By observing the wavelength dependence of both the magnitude of polarization (polarization fraction) and the polarization angle, we can characterize the dust grain properties and the physical conditions in a cloud. As discussed below, in the M17 cloud we find a compression shock front that is viewed edge-on. Thus this cloud provides a unique opportunity to study the polarization spectrum in regions having an anisotropic radiation field that varies spatially. This allows an experimental test of the RAT theory of grain alignment. M17 is also known as the Omega Nebula and is located in the constellation Sagittarius. This cloud is a premier example of a young, massive star formation region in the Galaxy. It is one of the brightest infrared and thermal radio sources in the sky. Its distance has been measured to be 1.6 $\pm$ 0.3 kpc \citep{2001A&A...377..273N}, and it spans an area of about $11\arcmin \times 9\arcmin$ across the sky. 

A geometric model of M17 was presented by \cite{2007ApJ...658.1119P}. In the inner part of the nebula, a bright, photoionized region with a hollow conical shape surrounds a central star cluster. This region expands outward in several directions into adjacent molecular gas. There is a large, unobscured optical \ion{H}{2} region spreading into the low density medium at the eastern edge of the molecular cloud. X-ray observations \citep{2003ApJ...590..306D, 2003ApJ...593..874T} have shown that the interior of the \ion{H}{2} region is filled by hot (with T $\sim 10^6$ - $10^7$ K) gas that is flowing out to the east. \cite{2003ApJ...590..306D} noted that this region is too young to have produced a supernova remnant and interpreted the X-ray emission as hot gas filling a super bubble blown by the OB stellar winds. In the middle of the nebula, velocity studies have shown an ionized shell having a diameter of about 6 pc \citep{2003ApJ...590..306D}. Toward some portions at the border of the ionized region, warm and hot gases are truncated by a wall of dense, cold molecular material that includes the dense cores known as M17 Southwest (M17 SW) and M17 North (M17 N). These cores exhibit many signposts of ongoing massive star formation. \cite{1977Afz....13..569G} characterized members of the young stellar cluster NGC6618 that is responsible for radiatively exciting the nebula.

In Section~\ref{sec:obs} of this paper, we describe new 450 $\mu$m polarimetric observations obtained for M17. Section~\ref{sec:genres} shows that our results for the magnetic field orientation are in good agreement with those from previous observations in the far-IR and submillimeter. In Section~\ref{sec:aveps}, we discuss the far-IR/submillimeter polarization spectrum of M17. In Section~\ref{sec:magchange}, we analyze the change of magnetic field across the shock front and find a correlation between the polarization angle and the location along an axis orthogonal to the shock front. We also find that the $P_{450}$/$P_{350}$ polarization ratio appears to be correlated with the strength of the radiation field, as discussed in Section~\ref{sec:pschange}. We explain both of these correlations in terms of the effects of stars in the central star cluster.

\section{Observations}
\label{sec:obs}

The 450 $\mu$m polarimetric data presented here were collected using the SHARP instrument \citep{2008ApOpt..47..422L} at the Caltech Submillimeter Observatory (CSO). SHARP is a fore-optics module that converts the SHARC II bolometer camera \citep{2003SPIE.4855...73D} into a sensitive imaging polarimeter with a spatial resolution of $\sim 11\arcsec$ at 450 $\mu$m, and $9\arcsec$ at 350 $\mu$m. The function of the fore-optics is to split the incident radiation in a $55\arcsec \times 55\arcsec$ field of view (FOV) into two orthogonally polarized beams that are then reimaged onto 12 $\times$ 12 pixel ``subarrays" at opposite ends of the 32 $\times$ 12 pixel array in SHARC II. The polarization signal is modulated by a stepped rotating half-wave plate (HWP) located skyward of the polarization-splitting optics. The observations were obtained during three nights in July 2010. The total integration time was about 9 hours, with an average zenith opacity $\tau \approx 1.3$ at 450 $\mu$m.

\section{Results}
\label{sec:results}

\subsection{General Results}
\label{sec:genres}

\begin{figure}
\centering
\includegraphics[width=0.48\textwidth]{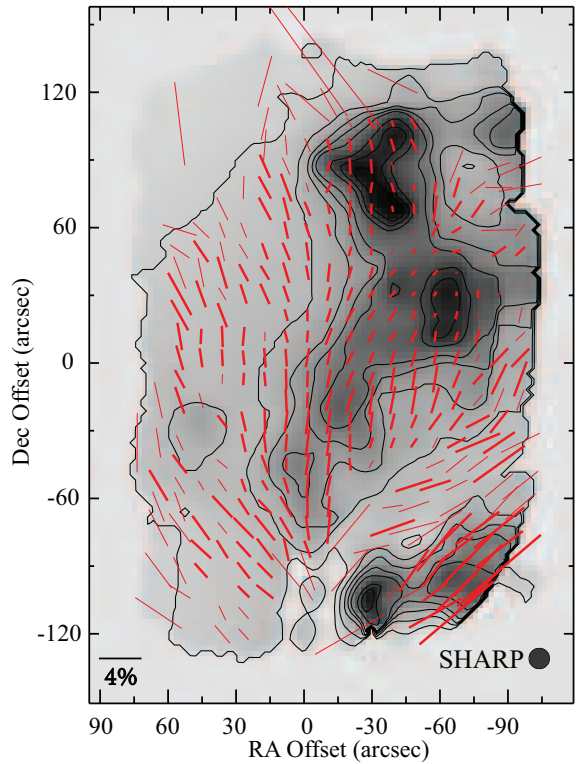}
\caption{M17 450 $\mu$m polarization vectors superposed on a map of the 450 $\mu$m intensity, also from SHARP data. Contours range from 10\% to 100\% of the peak intensity, in steps of 10\%. The three main flux peaks visible in this map correspond to the Northern, Central, and Southern Condensations of Figure 1b of \cite{1996ApJ...470..566D}. Thick vectors are detected with greater than or equal to $3\sigma$ sensitivity ($p \ge 3\sigma_{p}$) and thin vectors are between the $2\sigma$ and $3\sigma$ levels ($2\sigma_{p} \le p \le 3\sigma_{p}$). All vectors on the plot are corrected for positive bias. The orientation of each vector indicates the direction of the electric vector of the measured polarization. The key at bottom left shows the vector length corresponding to a polarization magnitude of 4\%. The circle on the bottom right shows the SHARP beam size. Right ascension and declination offsets are given with respect to $18^\text{h} 20^\text{m} 25^\text{s}$, $-16\arcdeg 13\arcmin 02\arcsec$ (J2000).}
\label{fig:SHARPpol}
\end{figure}

\begin{figure*}
\centering
\includegraphics[width=0.65\textwidth]{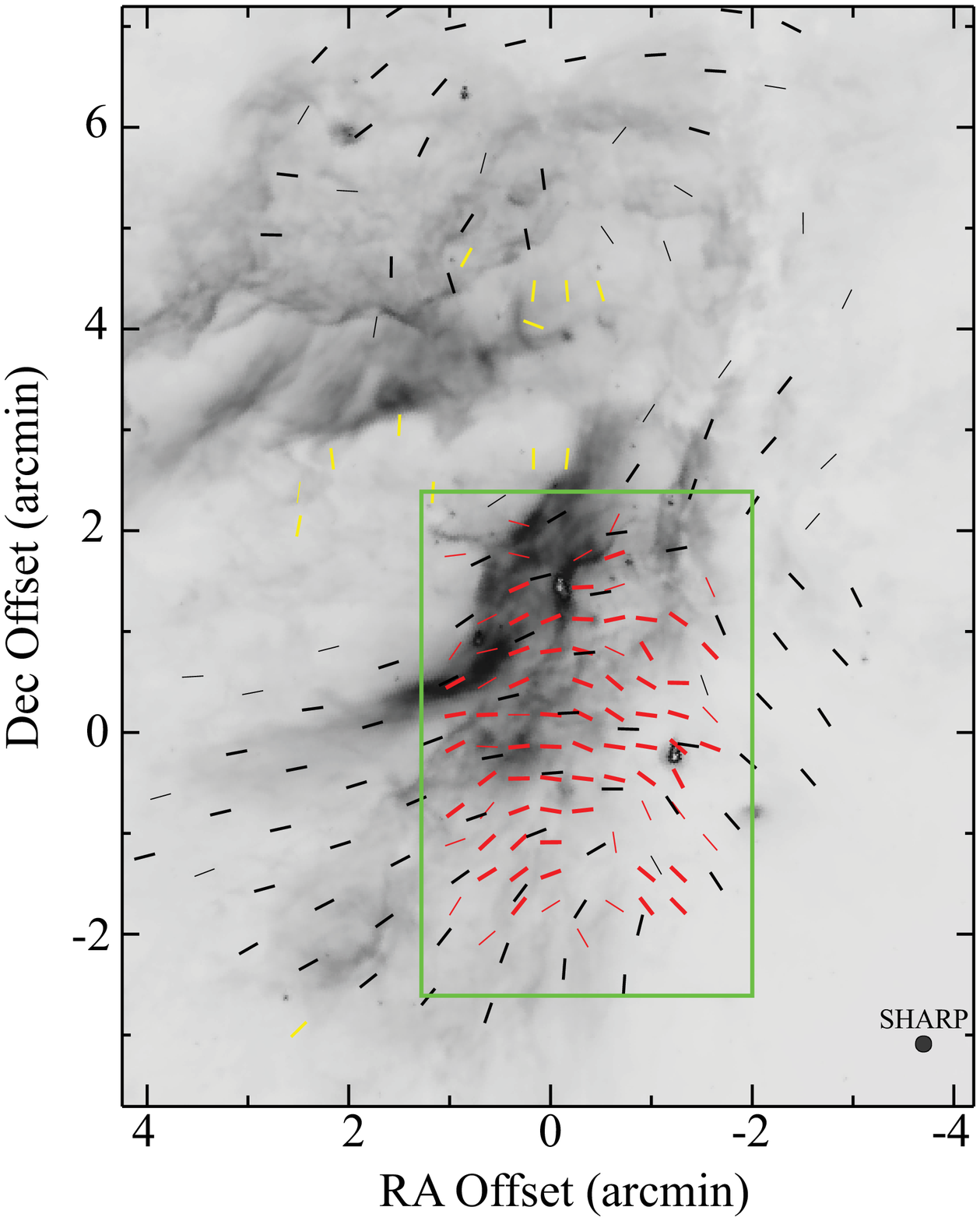}
\caption{Vectors show inferred magnetic field orientations from SHARP (red, 450 $\mu$m), Stokes (black, 100 $\mu$m; \citealp{1996ApJ...470..566D}) and optical observations (yellow; \citealp{1981A&A....95...94S}) superposed on a Spitzer/IRAC 8.0 $\mu$m image from GLIMPSE (Galactic Legacy Infrared Mid-Plane Survey Extraordinaire). The green box outlines the region displayed in Figure~\ref{fig:SHARPpol}. For clarity, not all vectors shown in Figure~\ref{fig:SHARPpol} are shown here. The hollow conical shape area in the center containing most of the yellow vectors is the \ion{H}{2} region. The cloud to the north of the \ion{H}{2} region is M17 N. The portion on the bottom right is M17 SW. The magnetic field orientations inferred from SHARP and Stokes data are perpendicular to the measured polarization angles, while those from optical polarization measurements are parallel to the polarization angles. Vectors shown here are plotted with uniform length and serve to indicate the inferred field orientation only. Thick vectors are detected with greater than or equal to $3\sigma$ sensitivity ($p \ge 3\sigma_{p}$) and thin vectors are between the $2\sigma$ and $3\sigma$ levels ($2\sigma_{p} \le p \le 3\sigma_{p}$). Right ascension and declination offsets are given with respect to $18^\text{h} 20^\text{m} 25^\text{s}$, $-16\arcdeg 13\arcmin 02\arcsec$ (J2000).}
\label{fig:Spitzer-optical}
\end{figure*}

Figure~\ref{fig:SHARPpol} shows our 450 $\mu$m polarization map of M17 superposed on contours of dust emission intensity, also taken from SHARP data. The map is centered at the J2000 coordinate ($18^\text{h} 20^\text{m} 25.1^\text{s}$, $-16^{\circ} 13' 02.1''$) and covers an area of about $4'25'' \times 2'45''$ overlapping M17 SW (see Figure~\ref{fig:Spitzer-optical}). Taking the distance to M17 to be 1.6 kpc, our map coverage corresponds to an area of 2.1 pc $\times$ 1.3 pc. The 450 $\mu$m M17 polarization measurements from SHARP are listed in Appendix~\ref{app:SHARPvector}. All the polarization magnitudes presented in this work, including the ones in figures and tables, are corrected for positive bias using the method described in \cite{2006PASP..118.1340V}.

As can be seen in Figure~\ref{fig:SHARPpol}, for regions of high submillimeter intensity, the average polarization fraction is lower than that for low intensity regions. This may be caused by any or all of the following effects: (1) If the magnetic field orientation varies along the line of sight (LOS), the measured polarization fraction becomes diluted upon integration along the LOS; (2) If the magnetic field within the dense part of the cloud is more ``tangled," averaging over the finite beam will cause a reduction in the observed polarization; (3) Grain alignment may be less efficient deep inside the dense part of the molecular cloud perhaps due to the weaker radiation field \citep{2005ApJ...631..361C, 2008ApJ...674..304W}.

For polarized emission, the magnetic field projected onto the plane of the sky is inferred by rotating the polarization E-vectors by $90\arcdeg$. Figure~\ref{fig:Spitzer-optical} shows the inferred magnetic field vectors from 100 $\mu$m \citep{1996ApJ...470..566D}, 450 $\mu$m (SHARP) and optical observations \citep{1981A&A....95...94S}. Our results for the magnetic field orientation are in good agreement with those from previous observations at far-IR wavelengths (Stokes instrument, 60 and 100 $\mu$m, $22\arcsec$ and $35\arcsec$ spatial resolution respectively; \citealp{1996ApJ...470..566D, 2000ApJS..128..335D}) and submillimeter wavelengths (Hertz instrument, 350 $\mu$m, $20\arcsec$ resolution; \citealp{2002ApJ...569..803H}), but have finer angular resolution. The 8.0 $\mu$m intensity from Spitzer GLIMPSE that is shown in the figure predominantly traces polycyclic aromatic hydrocarbon (PAH) molecular emission. The brightest 8.0 $\mu$m emission traces the boundary of the \ion{H}{2} region where the PAHs are being illuminated by the UV radiation from the central OB cluster, as seen in typical photon-dominated regions such as MonR2 and the Orion Bar \citep{2009ApJ...706L.160B, 2009A&A...498..161V}.

\clearpage

\cite{1996ApJ...470..566D} notes that the magnetic field revealed by the 100 $\mu$m polarization data bulges away from the \ion{H}{2} region (see black vectors in Figure~\ref{fig:Spitzer-optical}) and points out that this is consistent with the suggestion that the \ion{H}{2} region is expanding into its surroundings.  A similar situation was seen in the molecular cloud G333.6-0.2 by \cite{2006ApJ...648..340L}.  In this cloud, the authors found evidence of distortion of magnetic fields by an expanding photo-ionized gas bubble.  In the Galactic center, \cite{2000ApJ...529..241N} found that the expansion of the non-thermal shell source Sgr A (East) into a molecular cloud causes a similar effect and point out that, due to flux freezing, the magnetic field in an edge-on compression front should tend to run parallel to the compression front.  Indeed, this is approximately what is suggested by both the black \citep{1996ApJ...470..566D} and red (our work) vectors in Figure~\ref{fig:Spitzer-optical}, provided that we restrict ourselves to the boundary of the \ion{H}{2} region, where the 8.0 $\mu$m Spitzer GLIMPSE emission is strong.

\subsection{The Polarization Spectrum of M17}
\label{sec:aveps}

\begin{figure}[h!b]
\centering
\includegraphics[width=0.50\textwidth]{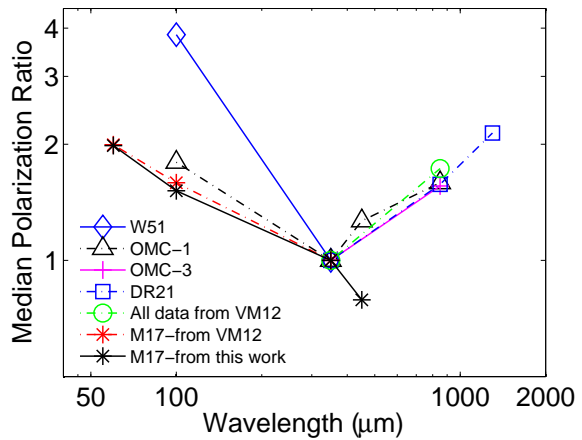}
\caption{Polarization spectrum of various interstellar molecular clouds from VM12 \citep{2012ApJS..201...13V} and this work. The green circle represents the median ratio for 15 clouds. The median polarization ratios are normalized to the value at 350 $\mu$m. Our results for the eastern part of M17 SW are in good agreement with the results of VM12 at 60, 100, and (by definition) 350 $\mu$m. In contrast to the results for OMC-1, our work shows that the eastern part of M17 SW has lower median polarization at 450 $\mu$m than at 350 $\mu$m. In this part of M17, the polarization spectrum falls monotonically from 60 $\mu$m to 450 $\mu$m.}
\label{fig:M17PS}
\end{figure}

Previous investigators have compared polarimetric data for various samples of molecular clouds at wavelengths ranging from the far-IR to submillimeter. If the source of the polarized emission is a single population of dust grains having identical polarization properties and temperature, the magnitude of the polarization (polarization fraction) is expected to be nearly independent of wavelength longward of 50 $\mu$m \citep{1988QJRAS..29..327H, 1999ApJ...516..834H}. However, the observed far-IR/submillimeter polarization spectra yield a different result, as shown in Figure~\ref{fig:M17PS} (see also \citealp{1999ApJ...516..834H, 2002ApJS..142...53V, 2004ASPC..309..515H, 2008ApJ...679L..25V, 2012ApJS..201...13V}). The polarization spectrum has been observed to fall from 60 to $\sim$ 350 $\mu$m (negative slope region) before rising again to 850 and 1300 $\mu$m (positive slope region), with its minimum located near 350 $\mu$m. In order to ensure that the computed polarization spectra are meaningful, several criteria are used when comparing the polarization fraction at two different wavelengths. For example, confusion can arise if the inclination of the field with respect to the LOS varies along the LOS. The likelihood of confusion due to this effect can be reduced by imposing the following criterion: The difference between the respective polarization angles at the two wavelengths, $|\Delta\phi|$, must be smaller than 10$\arcdeg$. This then leaves the alignment efficiency as a dominant factor affecting the polarization spectrum. This constraint is discussed in detail by \cite{2012ApJS..201...13V}.

Models containing two or more dust grain populations have been proposed to explain the observed structure in the polarization spectrum \citep{1999ApJ...516..834H, 2002ApJS..142...53V, 2007EAS....23..147V}. In such models, each dust grain population contributes a flux of $F_i(\nu) \propto \nu^{\beta_i} B_\nu(T_i)$, where $\nu$ is frequency, $B_\nu(T)$ is the Planck spectrum, and $\beta_i$ and $T_i$ are the spectral index and temperature of the dust population $i$, respectively. Correlation between the alignment efficiency and $\beta_i$ or $T_i$ for each dust population can result in a wavelength-dependent polarization spectrum. We will not refer to these early models in the remainder of this paper, focusing instead on more recent models that will be described below.

It has been pointed out that uncertainty could be introduced if one compares the polarization fraction ratio from two instruments having different chop throws, polarization efficiencies and/or beam sizes \citep{2012ApJS..201...13V}. The analysis presented here carefully combines the 450 $\mu$m SHARP polarimetric data ($\sim$ $11\arcsec$ resolution) with data collected at three other wavelengths. First, the 450 $\mu$m SHARP maps were smoothed to the same resolution as the 60 $\mu$m (Stokes, $22\arcsec$ resolution), 100 $\mu$m (Stokes, $35\arcsec$ resolution) and 350 $\mu$m (Hertz, $20\arcsec$ resolution) maps. Assuming that all beams were Gaussian, new maps of the Stokes parameters I, Q, and U were created from the original 450 $\mu$m maps by smoothing them with different Gaussian sizes to match the resolution of the 60, 100 and 350 $\mu$m data. Then, the polarization fractions and angles at the new resolution were calculated by resampling and combining the Stokes parameters in the new maps. 

At a given wavelength, the polarization vectors that are to be compared with vectors from the smoothed 450 $\mu$m map are chosen based on the following criteria adopted from \cite{1999ApJ...516..834H}: (1) The vectors are spatially separated by no more than 1$\arcsec$ in both RA and Dec; (2) The difference between the two polarization angles $|\Delta\phi|$ must be less than $10\arcdeg$; (3) The vectors are from the cloud envelope, not from high density cores; (4) All vectors are detected with signal to noise ratios greater than or equal to the $3 \sigma$ level.  

\begin{figure}[h!b]
\centering
\includegraphics[width=0.45\textwidth]{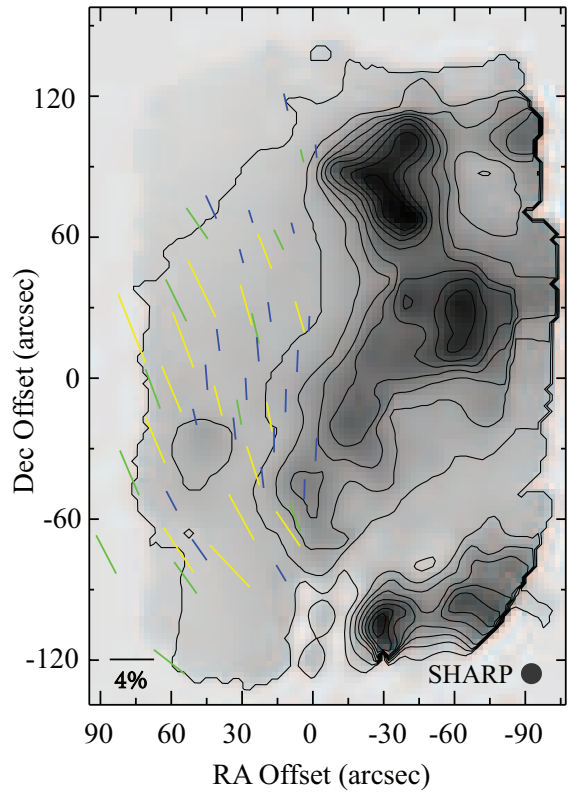}
\caption{Vectors selected for polarization spectrum analysis at 60 $\mu$m (yellow, Stokes, $22\arcsec$ resolution), 100 $\mu$m (green, Stokes, $35\arcsec$ resolution) and 350 $\mu$m (blue, Hertz, $20\arcsec$ resolution) superposed on the 450 $\mu$m intensity map from SHARP observations. Vectors are selected by comparing them with 450 $\mu$m data smoothed to matching angular resolution and then applying the selection criteria listed in Section~\ref{sec:aveps}. All vectors shown here are in the common region where we have data from all four wavelengths, i.e., the RA Offset $>$ 0 region (see data with $\Delta\alpha > 0$ in Tables~\ref{table:45060}, ~\ref{table:450100} and ~\ref{table:450350}). Right ascension and declination offsets are given with respect to $18^\text{h} 20^\text{m} 25^\text{s}$, $-16\arcdeg 13\arcmin 02\arcsec$ (J2000).}
\label{fig:M17PSarea}
\end{figure}

The M17 cloud spans a large area across the sky. Because we do not have polarimetric data for the entire cloud at all wavelengths, we can only compare the polarization ratios in a common region where we have data from all four wavelengths. This region is in the east portion of M17 SW with RA Offset $\geq$ 0 in Figure~\ref{fig:M17PSarea}. Polarization spectrum vectors for pairs of wavelengths (450 vs. 60, 450 vs. 60, and 450 vs. 350 $\mu$m) in the common region are listed in Tables~\ref{table:45060}, ~\ref{table:450100} and ~\ref{table:450350}. Table~\ref{table:450350} also lists the $P_{450}$/$P_{350}$ data for the RA Offset $<$ 0 region that will be discussed in Section~\ref{sec:pschange}. The polarization vectors plotted in Figure~\ref{fig:M17PSarea} are those from the RA Offset ($\Delta\alpha$) $>$ 0 region and are taken from Table~\ref{table:45060}, Table~\ref{table:450100} and  part of Table~\ref{table:450350}. The M17 polarization spectrum resulting from this work is calculated based on the data in the common region. It is superposed on previous spectra in Figure~\ref{fig:M17PS}, and tabulated in Table~\ref{table:450PS}. Our main result is that $P_{450}<P_{350}<P_{100}<P_{60}$ in the east portion of M17 SW. 

\clearpage

\begin{deluxetable}{ c c c c c c c c c c c c c}
\tablecaption{Polarization ratio data for 450 $\mu$m vs. 60 $\mu$m with 22$\arcsec$ resolution}
\tabletypesize{\footnotesize}
\tablecolumns{13}  
\tablewidth{0pt}
\tablehead{
		   \colhead{$\Delta\alpha$\,\tablenotemark{a}}         &
           \colhead{$\Delta\delta$\,\tablenotemark{a}}         &               
           \colhead{$P_{450}$}                                 &            
           \colhead{$\sigma_{p450}$}                              & 
           \colhead{P.A.\,\tablenotemark{b}}                   & 
           \colhead{$\sigma_\text{P.A.}$}                      & 
           \colhead{$I_{450}$\,\tablenotemark{c}}                   &
           \colhead{$P_{60}$}                                  &            
           \colhead{$\sigma_{p60}$}                              & 
           \colhead{P.A.\,\tablenotemark{b}}                   & 
           \colhead{$\sigma_\text{P.A.}$}                      & 
           \colhead{$I_{60}$\,\tablenotemark{c}}                   &
           \colhead{$P_{450}/P_{60}$\,\tablenotemark{d}}       \\
           \colhead{(arcsec)}                               &
           \colhead{(arcsec)}                               &
           \colhead{(\%)}                      			           &
           \colhead{(\%)}                           			   &
           \colhead{(deg)}                                &
           \colhead{(deg)}                                &
           \colhead{(-)}                      			           &
		   \colhead{(\%)}                      			           &
           \colhead{(\%)}                           			   &
           \colhead{(deg)}                                &
           \colhead{(deg)}                                &
           \colhead{(-)}                      			           &
           \colhead{(-)}                           			   }
\startdata
  80 &   19 &   2.1 &   0.5 &  18.5 &   5.9 &   0.26 &   6.7 &   0.5 &  21.7 &   2.2 &   0.47 &   0.32\\ 
  70 &  -29 &   1.5 &   0.3 &  21.9 &   5.2 &   0.34 &   4.5 &   0.3 &  23.0 &   1.6 &   0.66 &   0.33\\ 
  63 &   -7 &   1.8 &   0.2 &  17.6 &   3.8 &   0.30 &   4.5 &   0.2 &  22.8 &   1.4 &   0.71 &   0.40\\ 
  60 &  -76 &   2.5 &   0.4 &  38.2 &   4.2 &   0.24 &   4.8 &   0.5 &  33.6 &   2.8 &   0.32 &   0.52\\ 
  58 &   14 &   2.0 &   0.2 &  15.4 &   3.4 &   0.28 &   5.3 &   0.2 &  20.6 &   1.3 &   0.60 &   0.38\\ 
  50 &   36 &   2.4 &   0.3 &  23.4 &   3.0 &   0.25 &   5.7 &   0.3 &  26.0 &   1.7 &   0.56 &   0.42\\ 
  43 &  -12 &   1.1 &   0.2 &   7.3 &   4.3 &   0.32 &   2.8 &   0.2 &  14.3 &   1.8 &   0.83 &   0.39\\ 
  38 &  -83 &   2.1 &   0.2 &  40.6 &   2.7 &   0.26 &   5.2 &   0.4 &  43.5 &   2.1 &   0.31 &   0.40\\ 
  33 &  -62 &   1.4 &   0.2 &  33.0 &   3.0 &   0.33 &   4.7 &   0.3 &  28.8 &   1.7 &   0.45 &   0.30\\ 
  31 &   29 &   1.6 &   0.2 &  19.7 &   2.9 &   0.30 &   3.8 &   0.2 &  15.8 &   1.9 &   0.73 &   0.42\\ 
  28 &  -40 &   1.4 &   0.1 &  12.7 &   2.6 &   0.36 &   3.7 &   0.2 &  17.9 &   1.6 &   0.67 &   0.38\\ 
  23 &   52 &   1.6 &   0.2 &  20.4 &   3.2 &   0.28 &   3.4 &   0.5 &  21.3 &   3.7 &   0.75 &   0.47\\ 
  21 &  -19 &   1.5 &   0.1 &   1.8 &   2.1 &   0.37 &   2.7 &   0.2 &  10.2 &   1.7 &   0.88 &   0.56\\ 
  13 &  -67 &   1.5 &   0.1 &  26.4 &   2.2 &   0.43 &   3.8 &   0.3 &  34.4 &   2.0 &   0.39 &   0.40\\ 
   8 &   24 &   1.2 &   0.1 &  10.6 &   3.0 &   0.34 &   2.9 &   0.2 &  16.4 &   2.0 &   0.92 &   0.41\\ 
\enddata
\tablenotetext{a}{Offsets are given with respect to $18^\text{h} 20^\text{m} 25^\text{s}$, $-16\arcdeg 13\arcmin 02\arcsec$ (J2000).} 
\tablenotetext{b}{Position angle of electric vector measured east from north.}
\tablenotetext{c}{Intensity normalized to 1.00 at peak.}
\tablenotetext{d}{Median = 0.398, mean = 0.405 and std = 0.067 (see Table~\ref{table:450PS}).}
\label{table:45060}
\end{deluxetable}

%----------------------------------------------------%
\begin{deluxetable}{ c c c c c c c c c c c c c}
\tablecaption{Polarization ratio data for 450 $\mu$m vs. 100 $\mu$m with 35$\arcsec$ resolution}
\tabletypesize{\footnotesize}
\tablecolumns{13}  
\tablewidth{0pt}
\tablehead{\colhead{$\Delta\alpha$\,\tablenotemark{a}}         &
           \colhead{$\Delta\delta$\,\tablenotemark{a}}         &               
           \colhead{$P_{450}$}                                 &            
           \colhead{$\sigma_{p450}$}                              & 
           \colhead{P.A.\,\tablenotemark{b}}                   & 
           \colhead{$\sigma_\text{P.A.}$}                      & 
           \colhead{$I_{450}$\,\tablenotemark{c}}                   &
           \colhead{$P_{100}$}                                  &            
           \colhead{$\sigma_{p100}$}                              & 
           \colhead{P.A.\,\tablenotemark{b}}                   & 
           \colhead{$\sigma_\text{P.A.}$}                      & 
           \colhead{$I_{100}$\,\tablenotemark{c}}                   &
           \colhead{$P_{450}/P_{100}$\,\tablenotemark{d}}      \\
           \colhead{(arcsec)}                               &
           \colhead{(arcsec)}                               &
           \colhead{(\%)}                      			           &
           \colhead{(\%)}                           			   &
           \colhead{(deg)}                                &
           \colhead{(deg)}                                &
           \colhead{(-)}                      			           &
		   \colhead{(\%)}                      			           &
           \colhead{(\%)}                           			   &
           \colhead{(deg)}                                &
           \colhead{(deg)}                                &
           \colhead{(-)}                      			           &
           \colhead{(-)}                           			   }
\startdata
  91 &  -78 &   3.0 &   0.5 &  36.9 &   4.7 &   0.25 &   3.9 &   0.3 &  27.5 &   2.3 &   0.21 &   0.76\\ 
  81 &  -43 &   1.4 &   0.2 &  30.7 &   4.4 &   0.36 &   4.4 &   0.2 &  23.0 &   1.6 &   0.26 &   0.32\\ 
  71 &   -7 &   1.6 &   0.2 &  16.1 &   3.0 &   0.33 &   3.8 &   0.3 &  20.4 &   2.2 &   0.29 &   0.42\\ 
  64 & -124 &   1.5 &   0.3 &  50.5 &   6.1 &   0.25 &   3.5 &   0.3 &  51.9 &   2.9 &   0.14 &   0.42\\ 
  61 &   31 &   2.2 &   0.2 &  22.0 &   2.4 &   0.28 &   4.2 &   0.3 &  26.3 &   2.0 &   0.24 &   0.52\\ 
  57 &  -88 &   2.0 &   0.2 &  41.4 &   2.7 &   0.27 &   3.5 &   0.2 &  35.5 &   1.9 &   0.18 &   0.57\\ 
  52 &   64 &   1.7 &   0.2 &  28.6 &   3.6 &   0.26 &   3.4 &   0.4 &  34.7 &   3.1 &   0.17 &   0.50\\ 
  34 &  -17 &   1.2 &   0.1 &   5.7 &   1.9 &   0.39 &   2.3 &   0.2 &  10.1 &   1.8 &   0.44 &   0.52\\ 
  27 &   19 &   1.3 &   0.1 &   9.8 &   1.9 &   0.35 &   2.8 &   0.2 &  13.2 &   2.4 &   0.39 &   0.46\\ 
  17 &   57 &   1.5 &   0.1 &  17.2 &   1.8 &   0.35 &   2.0 &   0.1 &  25.5 &   1.9 &   0.36 &   0.75\\ 
  10 &  -62 &   1.4 &   0.1 &  11.4 &   1.4 &   0.51 &   2.7 &   0.3 &  21.3 &   3.0 &   0.37 &   0.52\\ 
   7 &   93 &   1.1 &   0.1 &  10.7 &   2.6 &   0.43 &   1.2 &   0.2 &  13.9 &   4.8 &   0.41 &   0.93\\
\enddata
\tablenotetext{a}{Offsets are given with respect to $18^\text{h} 20^\text{m} 25^\text{s}$, $-16\arcdeg 13\arcmin 02\arcsec$ (J2000).} 
\tablenotetext{b}{Position angle of electric vector measured east from north.}
\tablenotetext{c}{Intensity normalized to 1.00 at peak.}
\tablenotetext{d}{Median = 0.521, mean = 0.558 and std = 0.165 (see Table~\ref{table:450PS}).}
\label{table:450100}
\end{deluxetable}
%----------------------------------------------------%

\begin{deluxetable}{ c c c c c c c c c c c c c}
\tablecaption{Polarization ratio data for 450 $\mu$m vs. 350 $\mu$m with 20$\arcsec$ resolution}
\tabletypesize{\footnotesize}
\tablecolumns{13}  
\tablewidth{0pt}
\tablehead{\colhead{$\Delta\alpha$\,\tablenotemark{a}}         &
           \colhead{$\Delta\delta$\,\tablenotemark{a}}         &               
           \colhead{$P_{450}$}                                 &            
           \colhead{$\sigma_{p450}$}                              & 
           \colhead{P.A.\,\tablenotemark{b}}                   & 
           \colhead{$\sigma_\text{P.A.}$}                      & 
           \colhead{$I_{450}$\,\tablenotemark{c}}                   &
           \colhead{$P_{350}$}                                  &            
           \colhead{$\sigma_{p350}$}                              & 
           \colhead{P.A.\,\tablenotemark{b}}                   & 
           \colhead{$\sigma_\text{P.A.}$}                      & 
           \colhead{$I_{350}$\,\tablenotemark{c}}                   &
           \colhead{$P_{450}/P_{350}$\,\tablenotemark{d}}      \\
           \colhead{(arcsec)}                               &
           \colhead{(arcsec)}                               &
           \colhead{(\%)}                      			           &
           \colhead{(\%)}                           			   &
           \colhead{(deg)}                                &
           \colhead{(deg)}                                &
           \colhead{(-)}                      			           &
		   \colhead{(\%)}                      			           &
           \colhead{(\%)}                           			   &
           \colhead{(deg)}                                &
           \colhead{(deg)}                                &
           \colhead{(-)}                      			           &
           \colhead{(-)}                           			   }
\startdata
  63 &  -55 &   1.6 &   0.3 &  35.4 &   5.4 &   0.30 &   2.0 &   0.2 &  27.8 &   3.5 &   0.26 &   0.79\\ 
  53 &  -19 &   1.1 &   0.2 &  20.4 &   4.9 &   0.34 &   1.5 &   0.1 &  13.1 &   2.6 &   0.31 &   0.72\\ 
  51 &  -76 &   2.1 &   0.3 &  40.2 &   4.2 &   0.25 &   2.3 &   0.3 &  35.1 &   3.7 &   0.21 &   0.91\\ 
  48 &   -2 &   1.4 &   0.2 &   5.9 &   3.7 &   0.29 &   2.3 &   0.1 &   4.5 &   1.7 &   0.29 &   0.60\\ 
  46 &   71 &   1.8 &   0.4 &  30.1 &   6.5 &   0.22 &   2.3 &   0.4 &  24.5 &   5.1 &   0.18 &   0.77\\ 
  43 &   14 &   1.6 &   0.2 &   9.7 &   3.3 &   0.29 &   2.0 &   0.2 &   7.3 &   2.9 &   0.28 &   0.80\\ 
  36 &  -24 &   1.1 &   0.2 &  15.3 &   4.1 &   0.32 &   2.0 &   0.1 &   5.9 &   1.2 &   0.37 &   0.54\\ 
  33 &   50 &   1.7 &   0.2 &  22.0 &   3.8 &   0.26 &   1.3 &   0.2 &  15.0 &   3.9 &   0.25 &   1.31\\ 
  31 &   -7 &   1.3 &   0.2 &   2.4 &   3.3 &   0.32 &   2.0 &   0.1 &   3.0 &   1.0 &   0.38 &   0.64\\ 
  29 &   67 &   1.7 &   0.3 &  21.9 &   4.3 &   0.26 &   1.2 &   0.2 &  17.2 &   4.6 &   0.25 &   1.41\\ 
  26 &   10 &   1.3 &   0.1 &   2.8 &   3.2 &   0.31 &   2.2 &   0.1 &   5.0 &   1.0 &   0.39 &   0.59\\ 
  24 &  -45 &   1.4 &   0.1 &  12.0 &   2.3 &   0.39 &   1.9 &   0.1 &   6.7 &   0.9 &   0.48 &   0.74\\ 
  21 &   26 &   1.3 &   0.1 &  17.7 &   3.2 &   0.30 &   1.8 &   0.1 &   7.9 &   1.3 &   0.36 &   0.72\\ 
  19 &  -29 &   1.8 &   0.1 &  -0.1 &   1.9 &   0.39 &   2.2 &   0.1 & 180.0 &   0.6 &   0.49 &   0.82\\ 
  16 &  -86 &   2.2 &   0.2 &  35.2 &   2.4 &   0.31 &   1.7 &   0.2 &  30.9 &   3.4 &   0.31 &   1.30\\ 
  14 &  116 &   2.0 &   0.7 &  11.8 &   9.0 &   0.21 &   1.6 &   0.2 &  12.4 &   4.3 &   0.21 &   1.25\\ 
  14 &  -12 &   1.5 &   0.1 &   1.3 &   2.0 &   0.41 &   2.1 &   0.1 & 178.7 &   0.7 &   0.58 &   0.71\\ 
  11 &   62 &   1.7 &   0.2 &  19.3 &   2.7 &   0.31 &   1.0 &   0.1 &  14.6 &   3.4 &   0.36 &   1.70\\ 
   9 &    5 &   1.4 &   0.1 &   0.5 &   2.2 &   0.39 &   2.0 &   0.0 & 177.7 &   0.6 &   0.56 &   0.70\\ 
   6 &  -50 &   1.7 &   0.1 &  -0.9 &   1.5 &   0.52 &   1.8 &   0.1 & 178.6 &   0.8 &   0.55 &   0.94\\ 
   4 &   21 &   1.1 &   0.1 &   4.6 &   3.1 &   0.35 &   1.4 &   0.1 & 177.3 &   1.1 &   0.52 &   0.78\\ 
   1 &   95 &   1.0 &   0.2 &  15.6 &   5.3 &   0.39 &   1.2 &   0.1 &   5.6 &   1.3 &   0.50 &   0.82\\ 
   1 &  -33 &   2.2 &   0.1 &  -7.8 &   1.1 &   0.55 &   2.1 &   0.1 & 177.1 &   0.6 &   0.66 &   1.05\\ 
  -4 &  -17 &   1.8 &   0.1 &  -9.7 &   1.1 &   0.60 &   2.0 &   0.0 & 173.2 &   0.5 &   0.75 &   0.90\\ 
  -6 &  -74 &   2.0 &   0.2 &   0.4 &   2.9 &   0.43 &   0.6 &   0.1 &   2.3 &   3.3 &   0.36 &   3.36\\ 
  -6 &   57 &   1.2 &   0.1 &  11.3 &   2.8 &   0.38 &   0.5 &   0.1 &   6.6 &   2.9 &   0.60 &   2.44\\ 
  -9 &   -0 &   1.4 &   0.1 & -15.1 &   1.4 &   0.58 &   1.6 &   0.0 & 167.3 &   0.6 &   0.78 &   0.87\\ 
 -11 &  -57 &   1.6 &   0.2 &  -8.5 &   2.7 &   0.44 &   1.1 &   0.1 & 173.4 &   1.7 &   0.43 &   1.45\\ 
 -11 &   74 &   1.1 &   0.1 &  -5.2 &   2.3 &   0.57 &   0.8 &   0.0 & 177.8 &   1.2 &   0.87 &   1.37\\ 
 -14 &   17 &   1.3 &   0.1 & -22.2 &   1.8 &   0.50 &   1.1 &   0.0 & 162.1 &   0.8 &   0.77 &   1.18\\ 
 -16 &   90 &   1.2 &   0.1 &   3.8 &   1.8 &   0.78 &   1.3 &   0.0 &   4.8 &   0.6 &   0.94 &   0.92\\ 
 -16 &  -38 &   2.0 &   0.1 &  -7.4 &   1.4 &   0.54 &   1.2 &   0.1 & 174.5 &   1.2 &   0.54 &   1.67\\ 
 -19 &   33 &   1.1 &   0.1 & -18.6 &   2.3 &   0.49 &   0.7 &   0.0 & 156.9 &   1.6 &   0.76 &   1.56\\ 
 -21 &  -21 &   2.1 &   0.1 & -13.5 &   1.0 &   0.61 &   1.7 &   0.0 & 171.4 &   0.6 &   0.69 &   1.23\\ 
 -24 &   50 &   0.8 &   0.1 &  -6.5 &   2.7 &   0.59 &   0.4 &   0.0 & 174.3 &   3.7 &   0.84 &   1.98\\ 
 -26 &   -5 &   1.5 &   0.1 & -19.8 &   1.3 &   0.66 &   1.4 &   0.0 & 167.1 &   0.7 &   0.79 &   1.07\\ 
 -31 &   12 &   1.2 &   0.1 & -28.6 &   1.3 &   0.71 &   0.9 &   0.0 & 161.1 &   1.1 &   0.93 &   1.33\\ 
 -33 &  -45 &   0.8 &   0.1 & -20.1 &   4.5 &   0.46 &   0.8 &   0.1 & 153.8 &   2.1 &   0.43 &   1.00\\ 
 -33 &   86 &   1.0 &   0.1 &   5.9 &   1.7 &   0.98 &   0.9 &   0.0 &   9.4 &   0.9 &   1.00 &   1.11\\ 
 -38 &  102 &   1.0 &   0.1 &  20.0 &   2.5 &   0.88 &   1.3 &   0.0 &  11.7 &   0.7 &   0.81 &   0.77\\ 
 -38 &  -26 &   1.8 &   0.1 & -13.2 &   1.8 &   0.45 &   1.1 &   0.1 & 159.0 &   1.8 &   0.55 &   1.64\\ 
 -43 &   -10 &   1.9 &   0.1 & -12.9 &   1.2 &   0.55 &   1.4 &   0.1 & 165.7 &   1.3 &   0.67 &   1.36\\ 
 -46 &   64 &   0.6 &   0.1 &  -5.7 &   2.8 &   0.81 &   0.4 &   0.0 &   2.3 &   2.5 &   0.89 &   1.48\\ 
 -51 &   81 &   0.7 &   0.1 &  -8.5 &   4.0 &   0.74 &   0.6 &   0.0 & 179.1 &   2.0 &   0.67 &   1.15\\ 
 -53 & -105 &   3.4 &   0.3 & -46.1 &   2.6 &   0.56 &   2.1 &   0.2 & 141.6 &   2.3 &   0.38 &   1.62\\ 
 -56 &  -31 &   1.5 &   0.1 & -27.3 &   2.7 &   0.39 &   1.4 &   0.1 & 143.0 &   2.0 &   0.50 &   1.07\\ 
 -58 &  -88 &   3.8 &   0.4 & -43.5 &   2.4 &   0.46 &   2.5 &   0.1 & 140.7 &   1.4 &   0.51 &   1.51\\ 
 -66 &    2 &   1.2 &   0.1 & -16.6 &   1.3 &   0.75 &   0.9 &   0.1 & 164.5 &   2.5 &   0.78 &   1.34\\ 
 -68 &   76 &   1.5 &   0.2 & -26.3 &   3.4 &   0.43 &   0.5 &   0.1 & 151.3 &   3.1 &   0.59 &   3.03\\ 
 -71 &   19 &   0.7 &   0.0 & -11.9 &   1.8 &   0.94 &   0.7 &   0.1 & 177.0 &   3.1 &   0.91 &   1.01\\ 
 -71 & -109 &   4.8 &   0.4 & -47.7 &   2.2 &   0.63 &   2.8 &   0.3 & 136.5 &   3.4 &   0.24 &   1.72\\ 
 -73 &   93 &   0.7 &   0.3 & -21.7 &   9.4 &   0.45 &   0.5 &   0.1 & 161.5 &   3.1 &   0.68 &   1.51\\ 
 -75 &  -93 &   4.5 &   0.3 & -45.5 &   1.9 &   0.63 &   3.7 &   0.2 & 136.9 &   1.3 &   0.38 &   1.22\\ 
 -75 &  -19 &   1.6 &   0.1 & -29.3 &   2.5 &   0.45 &   1.5 &   0.1 & 147.7 &   1.9 &   0.51 &   1.07\\ 
 -80 &   55 &   0.7 &   0.1 & -55.4 &   5.0 &   0.52 &   0.2 &   0.1 & 123.4 &  10.8 &   0.72 &   4.00\\ 
 -85 &   71 &   1.4 &   0.2 & -49.6 &   4.9 &   0.40 &   0.9 &   0.1 & 128.6 &   1.8 &   0.68 &   1.55\\ 
 -90 &   88 &   1.4 &   0.4 & -63.9 &   7.4 &   0.46 &   0.6 &   0.1 & 119.9 &   2.3 &   0.76 &   2.44\\ 
 -98 &  -81 &   6.2 &   0.9 & -49.8 &   3.3 &   0.41 &   3.3 &   0.3 & 135.8 &   2.9 &   0.27 &   1.90\\ 
-103 &   67 &   1.7 &   0.4 & -52.9 &   6.9 &   0.50 &   0.8 &   0.2 & 128.3 &   7.6 &   0.56 &   2.13\\ 
-103 &  -64 &   4.3 &   1.1 & -50.4 &   6.5 &   0.33 &   3.6 &   0.3 & 139.3 &   2.3 &   0.33 &   1.19\\ 
-103 &   10 &   1.3 &   0.3 & -39.6 &   8.1 &   0.43 &   1.8 &   0.5 & 134.1 &   7.1 &   0.60 &   0.69\\ 
\enddata
\tablenotetext{a}{Offsets are given with respect to $18^\text{h} 20^\text{m} 25^\text{s}$, $-16\arcdeg 13\arcmin 02\arcsec$ (J2000).} 
\tablenotetext{b}{Position angle of electric vector measured east from north.}
\tablenotetext{c}{Intensity normalized to 1.00 at peak.}
\tablenotetext{d}{For vectors with $\Delta\alpha >= 0$, median = 0.790, mean = 0.897 and std = 0.294 (see Table~\ref{table:450PS}).}
\label{table:450350}
\end{deluxetable}

\begin{deluxetable}{c c c c c c}
\tablecaption{M17 Polarization Spectrum Data}
\tablecolumns{6}  
\tablewidth{0pt}
\tablehead{\colhead{Ratio}            &              
           \colhead{\# of Points}     &            
           \colhead{Median}           & 
           \colhead{Mean}             & 
           \colhead{Std}              & 
           \colhead{Ref.}             }
\startdata
$P_{450}/P_{60}$  & 15 & 0.398 & 0.405 & 0.067 & Table~\ref{table:45060}\\
$P_{450}/P_{100}$  & 12 & 0.521 & 0.558 & 0.165 & Table~\ref{table:450100}\\
$P_{450}/P_{350}$  & 23 & 0.790 & 0.897 & 0.294 & Table~\ref{table:450350}\\
\enddata
\label{table:450PS}
\end{deluxetable}
\clearpage

Past experience with other clouds shows that at a fixed wavelength, SHARP tends to produce higher polarization magnitudes than Hertz, even after smoothing to the same angular resolution (unpublished result by JEV). Our present work shows that M17 has lower median polarization at 450 $\mu$m (SHARP data) relative to that at 350 $\mu$m (Hertz data), for the east side of our map. If the above-mentioned offset between SHARP and Hertz is present in our data, then the actual median $P_{450}$/$P_{350}$ polarization ratio must be even lower than reported here. Thus, we are confident that the observed monotonic decrease in polarization ratio from 60 to 450 $\mu$m is a robust result. This suggests that for M17 the negative slope region extends beyond 350 $\mu$m, at least to 450 $\mu$m, which is different from what is seen in OMC-1 (Figure~\ref{fig:M17PS}).

According to the RAT theory of grain alignment, the alignment efficiency should be a function of grain environment. Simulations of a molecular cloud as a mixture of aspherical graphite and silicate under the RAT model results in a polarization spectrum rising from 100 to 450 $\mu$m and flat at longer wavelengths \citep{2007ApJ...663.1055B}. \cite{2009ApJ...696....1D} present similar models composed of both aspherical and spherical grains of both silicate and graphite composition. \cite{2009ApJ...696....1D} obtained their observational constraints on grain alignment from the optical to near-infrared polarization spectrum, and they obtained a monotonic polarization spectrum with a positive slope, even longward of 450 $\mu$m. 

\begin{figure}[h!b]
\centering
\includegraphics[width=0.45\textwidth]{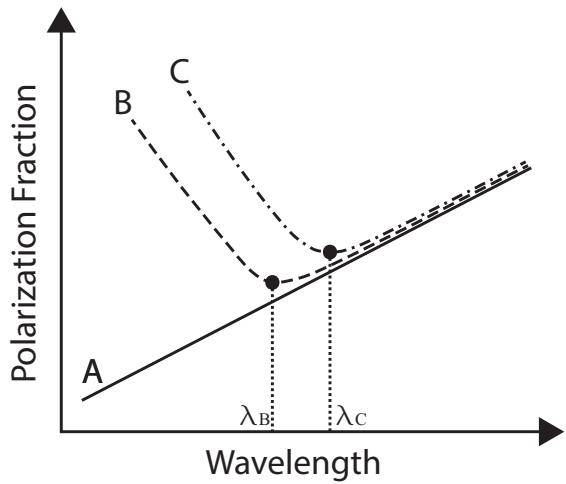}
\caption{Qualitative picture for how the cloud polarization spectrum might be expected to change as the power emitted by young stars in the cloud increases.  The solid curve (A) represents a cloud with no internal sources. The dashed (B) and dot-dashed (C) curves show the expectation for clouds containing, respectively, less powerful and more powerful internal sources. $\lambda_B$ and $\lambda_C$ are the locations of the respective polarization spectrum minima.}
\label{fig:RATalign}
\end{figure}

Figure~\ref{fig:RATalign} qualitatively illustrates our hypotheses concerning the polarization spectra for different clouds shown in Figure~\ref{fig:M17PS} in the context of the RAT alignment theory. To understand this, note that the two models described in the previous paragraph do not include internal radiation sources. These two models yield a monotonically increasing (on average) polarization spectrum, i.e. positive slope (see curve A of Figure~\ref{fig:RATalign}); they cannot account for the negative slope region of the polarization spectrum, which seems to be generally dominant between about 60 $\mu$m and 350 $\mu$m, corresponding roughly to the far-IR band (Figure~\ref{fig:M17PS}). However, note that the cool (10 -- 20 K) dust grain population that is considered in these models cannot explain the generally quite high intensity levels observed in the far-IR.  The far-IR emission from molecular clouds must instead be primarily due to dust heated to much higher temperatures by the intense radiation field created by embedded young stellar objects (YSOs) and young stars.  This warmer, highly irradiated dust would be expected to be very well aligned if grains are indeed aligned by the RAT mechanism.  Furthermore, the very hottest dust components should be the best aligned, which is precisely the recipe for a negative slope polarization spectrum. The observed far-IR/submillimeter spectra of Figure~\ref{fig:M17PS} may thus roughly be explained by drawing two component curves, a negative slope curve in the far-IR (warm dust irradiated by internal sources) and a positive slope curve at the longer submillimeter wavelengths (cool dust far from internal sources of radiation).  The position of the minimum is crudely set by the intersection of these two component curves (see Figure~\ref{fig:RATalign}).

\begin{figure*}
\centering
\includegraphics[width=0.65\textwidth]{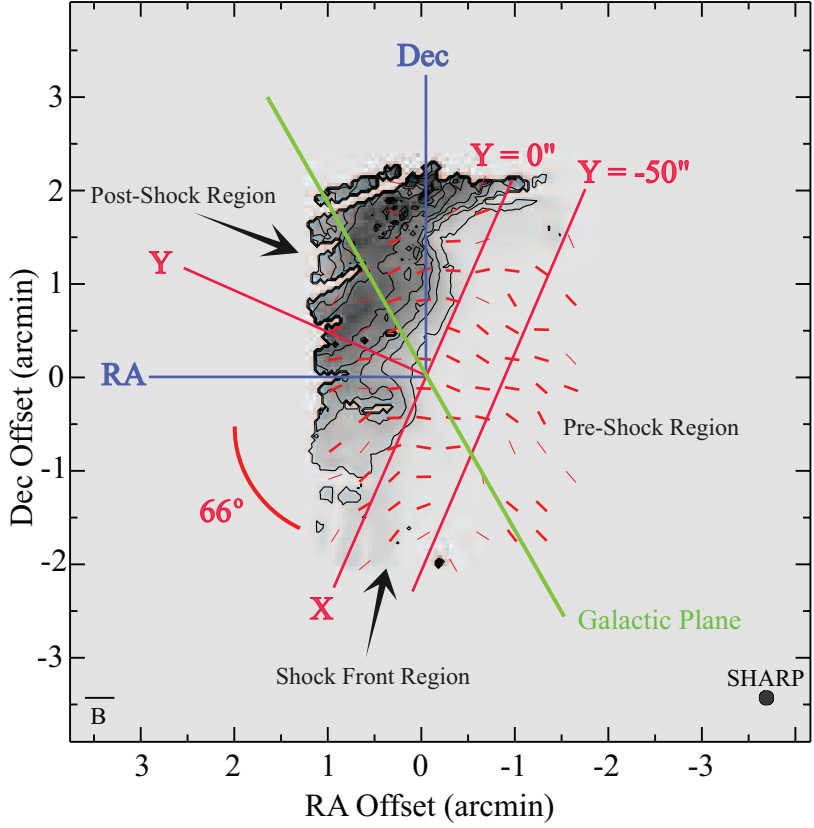}
\caption{Magnetic field vectors from SHARP plotted over the (21 cm)/(450 $\mu$m) intensity ratio map in arbitrary units (gray scale and contours), illustrating the compression shock front that is passing through the cloud. The contours range from 10\% to 100\% of the peak intensity ratio, in steps of 10\%. A new X-Y coordinate system is rotated by about $66\arcdeg$ with respect to the RA-Dec coordinates. The X axis aligns with the 10\% contour level. The shock proceeds the -Y direction. The Y = $0\arcsec$ and Y = -$50\arcsec$ lines separate the cloud into post-shock (Y $>$ $0\arcsec$), shock front (-$50\arcsec$ $<$ Y $<$ $0\arcsec$) and pre-shock (Y $<$ -$50\arcsec$) regions, as determined from variations in inferred magnetic field orientation (see Figure~\ref{fig:phiy}). For clarity, not all vectors from Figure~\ref{fig:SHARPpol} are shown here. Right ascension and declination offsets are given with respect to $18^\text{h}20^\text{m}25^\text{s}$, -$16\arcdeg13\arcmin02\arcsec$ (J2000).}
\label{fig:rotation}
\end{figure*}

In this picture, clouds with no internal sources would be expected to have only one component curve and thus only a positive slope region, shown as curve A in Figure~\ref{fig:RATalign}.  (This conjecture cannot be tested with present data, as no such clouds are included in the sample shown in Figure~\ref{fig:M17PS} because such quiescent clouds are faint shortward of 800 $\mu$m and have not been observed polarimetrically at these shorter wavelengths.)  If we include a few internal sources in a cloud, then we expect the negative slope component curve to be evident at least at the very shortest wavelengths where the contribution from cool dust far from radiation sources is negligible.  This is shown in Figure~\ref{fig:RATalign} as curve B.  Adding still more internal sources might be expected to increase the influence of the negative slope component curve, thus yielding curve C which has its minimum shifted to longer wavelengths.  In this interpretation, the fact that the minimum for the eastern part of M17 SW is shifted to the right with respect to that of OMC-1, i.e. from 350 $\mu$m to 450 $\mu$m or beyond (see Figure~\ref{fig:M17PS}), would indicate the existence of a stronger internal radiation field in the eastern portion of M17 SW in comparison with that in OMC-1.

\subsection{Changes in Magnetic Field Direction across the Shock Front}
\label{sec:magchange}

In Section~\ref{sec:intrm17} we reviewed how the central OB stars in M17 heat the \ion{H}{2} region and carve a hollow conical shape into the surrounding molecular cloud, and we noted that the M17 SW region provides a nearly edge-on view of the corresponding shock front. We can trace the progress of this shock front across M17 SW by studying the ratio of atomic column density to total column density. This ratio will be higher for the post-shock region. Figure~\ref{fig:rotation} shows 450 $\mu$m inferred magnetic field vectors superposed on a map of the ratio of 21 cm line intensity (VLA, \citealp{2001ApJ...560..821B}) to 450 $\mu$m continuum intensity (SHARP). This ``intensity ratio'' is a reasonable proxy for the ratio of atomic column density to total column density. We now define a new X-Y coordinate system for which the X axis is approximately coincident with the 10\% contour level of the normalized intensity ratio, as shown in Figure~\ref{fig:rotation}. This coordinate system is rotated counter-clockwise by an angle of about 66$\arcdeg$ with respect to the RA-Dec system. The shock front proceeds along the -Y direction.

\begin{figure}[h!b]
\centering
\includegraphics[width=0.5\textwidth]{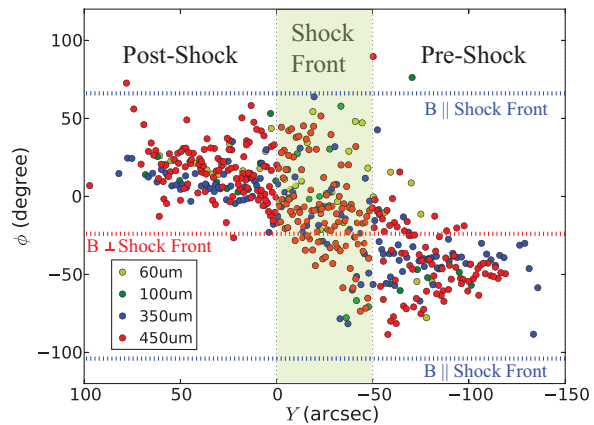}
\caption{Polarization angles as a function of perpendicular distance Y from post-shock/shock-front boundary line (see Figure~\ref{fig:rotation}). The post-shock region has Y $> 0\arcsec$ and the pre-shock region has Y $< -50\arcsec$. The 60 $\mu$m data points are in yellow, 100 $\mu$m data in green, 350 $\mu$m data in blue, and 450 $\mu$m data in red. We only show data from the region within $-120\arcsec <$ RA Offset $< 70\arcsec$ and $-120\arcsec <$ Dec Offset $< 120\arcsec$ (see Figure~\ref{fig:rotation}).}
\label{fig:phiy}
\end{figure}

In Figure~\ref{fig:phiy}, the polarization angle for multi-wavelength data toward M17 SW is shown as a function of the Y coordinate defined above. We can clearly see a strong correlation between the two quantities. Based on the variation of polarization angle seen in Figure~\ref{fig:phiy}, we use two lines, defined by Y = 0$\arcsec$ and Y = -50$\arcsec$, to separate the cloud into three regions: post-shock, shock front and pre-shock. These three regions are indicated in both Figure~\ref{fig:rotation} and Figure~\ref{fig:phiy}.

In Section~\ref{sec:genres} we reviewed the arguments presented by \cite{1996ApJ...470..566D} and others regarding the manner in which a compression front or shock front is expected to distort a cloud's magnetic field. Specifically, the effect of the compression should be to force the magnetic field nearly parallel to the observed compression front. Figure~\ref{fig:phiy} gives the polarization angle values corresponding to magnetic field parallel to and perpendicular to the shock front, which are respectively at 66$\arcdeg$ and -24$\arcdeg$. In order to quantitatively explore the effects of the shock, we computed mean polarization angles for the pre-shock, shock front, and post-shock regions. Due to the 180$\arcdeg$ ambiguity in polarization angle, the mean angle is, strictly speaking, not well defined. The equal weight Stokes mean (EWSM) technique \citep{2006ApJ...648..340L} provides a simple and unambiguous method for computing an effective mean angle, and using this technique we find mean polarization angles of -46$\arcdeg$, 0$\arcdeg$, and 18$\arcdeg$, respectively, for the pre-shock, shock front, and post-shock regions. From the above values, we can see that the pre-shock region has its inferred field rotated by 70$\arcdeg$ counter-clockwise with respect to the shock front, while for the post-shock region the field is 48$\arcdeg$ clockwise from the shock front. It appears that, as expected, the shock front has changed the field direction in the sense of making it lie closer to parallelism with the shock front (see also Figure~\ref{fig:phiy}). However, the angle between the post-shock magnetic field and the shock front is still substantial, for reasons that are not clear.

The magnetic field in the pre-shock region of M17 SW may be taken as representative of the unperturbed field of the M17 molecular cloud. This field has an average direction of 44$\arcdeg$, and thus lies within 17$\arcdeg$ of the Galactic plane which is at position angle 27$\arcdeg$ (see Figure~\ref{fig:rotation}). This result is consistent with the conclusion by \cite{2006ApJ...648..340L} that Giant Molecular Cloud magnetic fields run preferentially parallel to the Galactic plane.

We have argued that we can understand gross features of the magnetic field in M17 SW in terms of simple arguments previously advanced by \cite{1996ApJ...470..566D}, \cite{2000ApJ...529..241N}, and \cite{2006ApJ...648..340L}. Our two-dimensional analysis of cloud structure and polarization reveals some physical processes in the shock front region, which has an edge-on shell structure. However, we still lack a full understanding of the magnetic field in this source. In particular, both the magnetic field and the shell structure are inherently three-dimensional. A full three-dimensional model including the results of Zeeman mapping of the line-of-sight field (e.g., \citealp{2001ApJ...560..821B}) might be helpful, but is beyond the scope of the present paper.

\subsection{Changes in Polarization Spectrum across the Shock Front}
\label{sec:pschange}

In warmer regions with higher radiation fields, the dust grains would be expected to be better aligned, if RAT theory is correct. Statistically higher polarization fractions are observed in warmer regions \citep{2012ApJS..201...13V}. We cannot observe this easily in M17 since the magnetic field inclination with respect to the LOS likely varies in a complex manner, due to the shock front in M17 \citep{1999ApJ...515..304B}. However, this geometric effect should not affect the polarization spectrum, due to our use of the $|\Delta\phi| < 10\arcdeg$ criterion (see Section~\ref{sec:aveps}). Thus, we turn to the issue of polarization spectrum variations across the shock front.

\begin{figure}[h!b]
\centering
\includegraphics[width=0.48\textwidth]{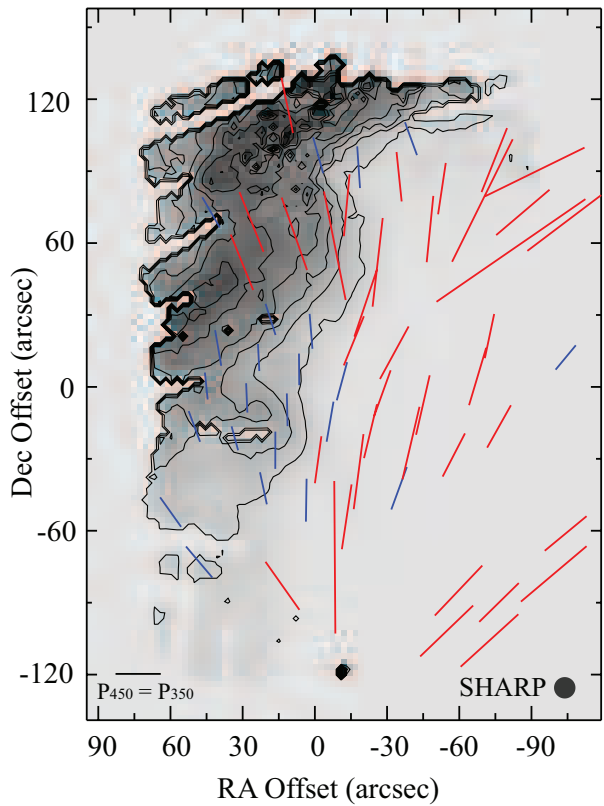}
\caption{The $P_{450}$/$P_{350}$ polarization ratio vectors superposed on the (21 cm)/(450 $\mu$m) intensity ratio map from Figure~\ref{fig:rotation}. The length of each vector is proportional to the ratio of $P_{450}/P_{350}$. The orientations of the vectors are parallel to the polarization angles of the 450 $\mu$m data. The blue vectors represent $P_{450}$ $<$ $P_{350}$, while the red vectors correspond to $P_{450}$ $>$ $P_{350}$. The scale at bottom left is equivalent to $P_{450}/P_{350} = 1.0$. Right ascension and declination offsets are given with respect to $18^\text{h} 20^\text{m} 25^\text{s}$, $-16\arcdeg 13\arcmin 02\arcsec$ (J2000).}
\label{fig:21cm-M17}
\end{figure}

\begin{figure}[h!b]
\centering
\includegraphics[width=0.50\textwidth]{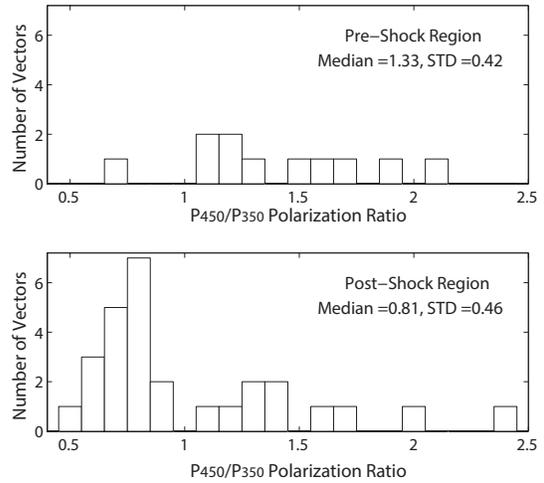}
\caption{Histograms of $P_{450}$/$P_{350}$ polarization ratio for pre-shock and post-shock regions. The median ratios of these two spatial regions are respectively 1.33 and 0.81, with standard deviations of 0.42 and 0.46. More measurements of polarization ratio corresponding to negative slope are found in the post-shock region, and more positive slope ratios are found in the pre-shock region.}
\label{fig:Hist}
\end{figure}

The average polarization spectrum that was plotted in Figure~\ref{fig:M17PS} is dominated by vectors in the post-shock region due to limitations of the spatial extent of the 60 and 100 $\mu$m data. In this region $P_{450}$ is smaller than $P_{350}$. However, in the shock front and pre-shock regions, we find that the median of $P_{450}$ is greater than that of $P_{350}$. Using the data presented in Table~\ref{table:450350}, Figure~\ref{fig:21cm-M17} shows the $P_{450}$/$P_{350}$ polarization ratio vector superposed on the (21 cm)/(450 $\mu$m) intensity ratio map. Most of the blue vectors ($P_{450} < P_{350}$) are in the post-shock region. In spite of some red ($P_{450} > P_{350}$) vectors distributed around the densest part of the cloud, the contour level (21 cm)/(450 $\mu$m) = 0.1 roughly separates the respective areas for blue and red vectors. Figure~\ref{fig:Hist} shows separate histograms of $P_{450}$/$P_{350}$ polarization ratio for the pre-shock and post-shock regions, respectively. The median polarization ratios of these two spatial regions are 1.33 and 0.81, respectively. There are more measurements of polarization ratio corresponding to negative slope in the post-shock region and the opposite is true in the pre-shock region. If the RAT theory is correct, the stronger radiation field in the post-shock east part of the cloud should cause the minimum in the polarization spectrum to shift toward longer wavelength (see Figure~\ref{fig:RATalign}) for this region. Thus the $P_{450} / P_{350}$ ratio should become smaller in the post-shock region, which is exactly what we see in Figure~\ref{fig:Hist}.

\begin{figure}[h!b]
\centering
\includegraphics[width=0.48\textwidth]{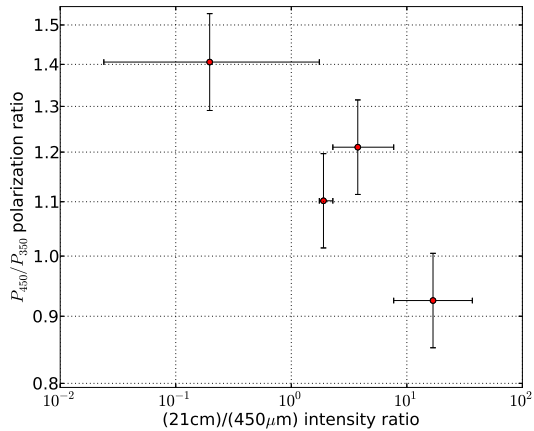}
\caption{The $P_{450}$/$P_{350}$ polarization ratio plotted against the (21 cm)/(450 $\mu$m) intensity ratio. The data used for this plot are the same as shown in Figure~\ref{fig:21cm-M17}. The bins for the data points are (0.024 -- 1.748), (1.748 -- 2.289), (2.289 -- 7.695) and (7.695 -- 36.877) in the (21 cm)/(450 $\mu$m) intensity ratio. For each bin, the data point and vertical error bar respectively represent the mean value of the $P_{450}/P_{350}$ ratio for that bin and the uncertainty of this mean value. The horizontal error bars show the ranges of the bins. These ranges were determined so as to keep the vertical error bars for the bins as close as possible to one another.}
\label{fig:binned}
\end{figure}

Figure~\ref{fig:binned} shows the correlation between the $P_{450}$/$P_{350}$ polarization ratio and the (21 cm)/(450 $\mu$m) intensity ratio. In this figure, data shown in Figure~\ref{fig:21cm-M17} are binned into four bins having the following (21 cm)/(450 $\mu$m) intensity ratios: (0.024 -- 1.748), (1.748 -- 2.289), (2.289 -- 7.695) and (7.695 -- 36.877). The bins sizes have been chosen to make the error bars in $P_{450}$/$P_{350}$ polarization ratio (vertical error bars) approximately equal. The data points and vertical error bars in Figure~\ref{fig:binned} show the mean value of the $P_{450}/P_{350}$ ratio for each bin and the uncertainty of this value, respectively. These uncertainties were calculated assuming Gaussian errors in the individual ratios shown in Figure~\ref{fig:21cm-M17}. (I.e., the plotted uncertainty is the r.m.s. of the ratios divided by the square root of the number of ratios used to compute the corresponding mean value). The bin sizes are represented by the horizontal error bars. Although the vertical error bars are large, we do see a trend of falling $P_{450}/P_{350}$ ratio, as we progress from molecular-dominated to atomic-dominated regions. The horizontal axis of Figure~\ref{fig:binned} provides a measurement of the strength of the radiation field. We see that the atomic-dominated regions, which have greater exposure to radiation sources in comparison with the molecular-dominated regions, exhibit a shift to a negative slope just as expected if the minimum is being pushed toward longer wavelength as illustrated in Figure~\ref{fig:RATalign}. Future observations covering more wavelengths could potentially reduce the error bars in the plot of Figure~\ref{fig:binned}, confirming the existence of this tentative correlation between dust grain environment and polarization spectrum.

\section{Summary}

The combination of general multi-wavelength studies and polarimetric data in the far-IR and submillimeter allows us to probe the physical structure and evolution of the M17 cloud. At large scales, young OB type stars in the center of the cloud heat the \ion{H}{2} region up to $10^6$ - $10^7$ K and create a high energy fountain towards the southeast direction. The \ion{H}{2} wind pushes the \ion{H}{1} and $\text{H}_2$ regions outward, creating a hollow region with a conical shape. The magnetic field is found bulging away from the \ion{H}{2} region, as noted earlier by \cite{1996ApJ...470..566D}. 

At small scales, within M17 SW, a compression shock front is found at the boundary between the \ion{H}{1} and $\text{H}_2$ regions. There are significant differences in the polarization spectrum between the pre-shock and post-shock regions. Specifically, the negative slope region is dominant (extending from 60 to 450 $\mu$m) in the east (post-shock) part of M17 SW where there is a stronger radiation field. In the west (pre-shock) part, where the radiation is weaker, the positive slope region begins to dominate by 350 $\mu$m. We have suggested that this change is in qualitative agreement with predictions of the radiative alignment torque (RAT) model for grain alignment. Grains in molecular clouds are not always aligned with the magnetic field perfectly, and any model trying to explain the polarization spectrum should take into account the variance of interstellar physical conditions across a cloud as well as along the line of sight. 

\acknowledgments
We thank C. Brogan for providing the 21 cm intensity data. We are grateful to A. Chepurnov, R. Hildebrand, and A. Lazarian for illuminating discussions. This material is based upon work at the Caltech Submillimeter Observatory, which is operated by the California Institute of Technology under cooperative agreement with the National Science Foundation (AST-0838261). We thank the National Science Foundation for supporting SHARP, via grant AST-0909030 to Northwestern University.

%---------------------------
\bibliographystyle{apj}
\bibliography{M17}
%---------------------------

\appendix
\section{450 $\mu$m polarization measurements for M17}
\label{app:SHARPvector}

\setcounter{table}{0}
\renewcommand*\thetable{\Alph{section}.\arabic{table}}

\begin{deluxetable}{c c c c c c c}
\tabletypesize{\footnotesize}
\tablecaption{450 $\mu$m polarization measurements for M17}
\tablecolumns{7}  
\tablewidth{0pt}
\tablehead{\colhead{$\Delta\alpha$\,\tablenotemark{a}}         &
           \colhead{$\Delta\delta$\,\tablenotemark{a}}         &               
           \colhead{P}                                         &            
           \colhead{$\sigma_{p}$}                              & 
           \colhead{P.A.\,\tablenotemark{b}}                   & 
           \colhead{$\sigma_\text{P.A.}$}                      & 
           \colhead{I\,\tablenotemark{c}}                   \\
           \colhead{(arcsec)}                               &
           \colhead{(arcsec)}                               &
           \colhead{(\%)}                      			           &
           \colhead{(\%)}                           			   &
           \colhead{(deg)}                               &
           \colhead{(deg)}                               &
           \colhead{(-)}                      			           }
\startdata
  80.0 &  -38.0 &   5.0 &   2.5 &  -0.2 &  11.7 &  0.25\\ 
  70.2 & -114.0 &   4.8 &   2.0 &  56.3 &  11.0 &  0.19\\ 
  70.2 &  -95.0 &   3.8 &   1.6 &  36.8 &  11.5 &  0.19\\ 
  70.2 &  -76.0 &   3.4 &   1.6 &  27.2 &  11.6 &  0.20\\ 
  70.2 &  -66.5 &   2.7 &   1.3 &  21.6 &  12.8 &  0.21\\ 
  70.1 &  -57.0 &   3.2 &   1.1 &  30.8 &   8.8 &  0.24\\ 
  70.1 &  -38.0 &   2.5 &   1.2 &   8.5 &  11.6 &  0.26\\ 
  70.1 &  -19.0 &   3.3 &   1.2 &  12.2 &   9.3 &  0.26\\ 
  60.3 & -104.5 &   2.1 &   1.1 &  58.2 &  12.8 &  0.21\\ 
  60.3 &  -95.0 &   2.6 &   1.0 &  25.4 &  10.6 &  0.20\\ 
  60.3 &  -76.0 &   3.2 &   1.2 &  36.1 &   9.7 &  0.20\\ 
  60.3 &  -66.5 &   2.6 &   1.3 &  19.6 &  13.1 &  0.22\\ 
  60.3 &  -57.0 &   1.7 &   0.7 &  30.7 &  11.5 &  0.26\\ 
  60.3 &  -47.5 &   2.1 &   0.6 &  36.9 &   7.5 &  0.29\\ 
  60.3 &  -38.0 &   1.4 &   0.6 &  24.1 &  11.6 &  0.30\\ 
  60.2 &  -19.0 &   2.6 &   0.6 &  18.9 &   6.3 &  0.28\\ 
  60.2 &   -9.5 &   1.8 &   0.6 &  27.6 &   8.1 &  0.27\\ 
  60.2 &   -0.0 &   2.3 &   0.7 &  15.7 &   8.2 &  0.25\\ 
  60.2 &    9.5 &   2.6 &   0.5 &   9.5 &   5.8 &  0.24\\ 
  60.2 &   19.0 &   3.2 &   0.8 &  29.3 &   6.8 &  0.23\\ 
  60.2 &   28.5 &   4.4 &   1.0 &  28.8 &   5.9 &  0.22\\ 
  60.2 &   38.0 &   3.1 &   1.1 &  20.6 &   9.6 &  0.21\\ 
  60.2 &   47.5 &   2.0 &   0.9 &  56.0 &  11.1 &  0.21\\ 
  60.2 &   57.0 &   3.0 &   1.4 &  72.6 &  11.3 &  0.20\\ 
  60.2 &  104.5 &   7.0 &   3.6 &   6.9 &  11.5 &  0.17\\ 
  50.4 &  -95.0 &   1.7 &   0.6 &  41.9 &   9.1 &  0.21\\ 
  50.4 &  -85.5 &   2.5 &   0.6 &  47.5 &   6.0 &  0.21\\ 
  50.4 &  -76.0 &   2.7 &   0.7 &  35.6 &   7.3 &  0.22\\ 
  50.4 &  -57.0 &   0.9 &   0.5 &  29.8 &  14.0 &  0.27\\ 
  50.4 &  -19.0 &   1.4 &   0.6 &  15.4 &  12.1 &  0.27\\ 
  50.4 &   -0.0 &   1.7 &   0.6 &   7.9 &   8.8 &  0.25\\ 
  50.4 &    9.5 &   1.3 &   0.4 &  -2.5 &   8.6 &  0.25\\ 
  50.4 &   28.5 &   2.6 &   0.6 &  19.9 &   6.2 &  0.23\\ 
  50.3 &   38.0 &   2.6 &   1.0 &  11.4 &   9.8 &  0.22\\ 
  50.3 &   47.5 &   2.3 &   0.7 &  34.2 &   8.5 &  0.21\\ 
  50.3 &   57.0 &   1.9 &   0.9 &  46.0 &  12.3 &  0.21\\ 
  40.5 & -123.5 &   2.0 &   0.9 &  40.3 &  11.6 &  0.21\\ 
  40.5 & -114.0 &   1.3 &   0.7 &  25.2 &  14.1 &  0.21\\ 
  40.5 &  -85.5 &   1.9 &   0.5 &  31.9 &   7.8 &  0.23\\ 
  40.5 &  -76.0 &   1.2 &   0.6 &  38.8 &  12.8 &  0.23\\ 
  40.5 &  -66.5 &   2.7 &   0.9 &  43.0 &   8.7 &  0.25\\ 
  40.5 &  -57.0 &   1.3 &   0.5 &  34.7 &   9.7 &  0.27\\ 
  40.5 &  -47.5 &   1.2 &   0.5 &  50.3 &   9.8 &  0.27\\ 
  40.5 &  -28.5 &   1.4 &   0.4 &  37.2 &   7.6 &  0.30\\ 
  40.5 &   -9.5 &   0.8 &   0.4 &  -2.6 &  13.4 &  0.26\\ 
  40.5 &   -0.0 &   1.5 &   0.5 &   3.0 &   8.9 &  0.26\\ 
  40.5 &    9.5 &   2.0 &   0.4 &   2.2 &   5.7 &  0.25\\ 
  40.5 &   19.0 &   2.3 &   0.6 &  21.0 &   6.4 &  0.25\\ 
  40.5 &   28.5 &   1.2 &   0.5 &  30.7 &  11.5 &  0.25\\ 
  40.5 &   38.0 &   3.0 &   0.7 &  16.5 &   5.5 &  0.23\\ 
  40.5 &   47.5 &   1.7 &   0.8 &  12.7 &  12.5 &  0.22\\ 
  40.5 &   57.0 &   1.3 &   0.7 &  20.6 &  13.0 &  0.21\\ 
  40.5 &   66.5 &   2.2 &   0.9 &  26.4 &  10.7 &  0.21\\ 
  30.6 & -123.5 &   1.2 &   0.7 &  42.1 &  13.4 &  0.24\\ 
  30.6 & -114.0 &   1.4 &   0.5 &  40.7 &  10.6 &  0.24\\ 
  30.6 & -104.5 &   1.4 &   0.5 &  45.2 &   9.0 &  0.24\\ 
  30.6 &  -95.0 &   1.9 &   0.4 &  37.8 &   6.4 &  0.24\\ 
  30.6 &  -85.5 &   2.0 &   0.4 &  42.1 &   6.1 &  0.25\\ 
  30.6 &  -76.0 &   2.7 &   0.5 &  44.3 &   5.3 &  0.25\\ 
  30.6 &  -66.5 &   1.7 &   0.5 &  33.0 &   8.3 &  0.27\\ 
  30.6 &  -57.0 &   1.2 &   0.4 &  19.0 &   9.0 &  0.28\\ 
  30.6 &  -38.0 &   1.7 &   0.4 &  12.3 &   6.4 &  0.27\\ 
  30.6 &  -28.5 &   1.8 &   0.5 &  12.8 &   6.8 &  0.27\\ 
  30.6 &   -9.5 &   1.1 &   0.4 &  -3.5 &   8.8 &  0.27\\ 
  30.6 &   -0.0 &   1.3 &   0.4 &   6.0 &   7.9 &  0.27\\ 
  30.6 &    9.5 &   1.8 &   0.4 &  -5.3 &   5.7 &  0.27\\ 
  30.6 &   28.5 &   1.8 &   0.6 &  18.0 &   8.4 &  0.25\\ 
  30.6 &   38.0 &   1.8 &   0.4 &  27.6 &   6.7 &  0.25\\ 
  30.6 &   47.5 &   1.5 &   0.5 &  30.9 &   9.9 &  0.24\\ 
  30.6 &   57.0 &   1.4 &   0.6 &  21.2 &  11.6 &  0.23\\ 
  30.6 &   76.0 &   2.0 &   1.0 &  -1.7 &  12.0 &  0.21\\ 
  20.7 & -104.5 &   2.0 &   0.5 &  51.5 &   7.4 &  0.25\\ 
  20.7 &  -95.0 &   2.0 &   0.4 &  44.9 &   6.4 &  0.26\\ 
  20.7 &  -85.5 &   2.5 &   0.4 &  36.7 &   3.9 &  0.28\\ 
  20.7 &  -76.0 &   3.0 &   0.5 &  38.5 &   4.2 &  0.28\\ 
  20.7 &  -66.5 &   1.8 &   0.4 &  43.3 &   6.4 &  0.29\\ 
  20.7 &  -57.0 &   1.1 &   0.3 &  29.5 &   7.1 &  0.34\\ 
  20.7 &  -47.5 &   1.8 &   0.3 &  20.1 &   4.9 &  0.35\\ 
  20.7 &  -38.0 &   1.9 &   0.4 &   6.3 &   5.2 &  0.30\\ 
  20.7 &  -28.5 &   2.4 &   0.4 &   1.1 &   4.5 &  0.29\\ 
  20.7 &   -9.5 &   1.4 &   0.3 &   5.4 &   6.7 &  0.29\\ 
  20.7 &   -0.0 &   1.9 &   0.3 &   4.6 &   5.0 &  0.29\\ 
  20.7 &    9.5 &   0.8 &   0.4 &  -0.1 &  12.4 &  0.28\\ 
  20.7 &   28.5 &   2.0 &   0.5 &  24.3 &   6.8 &  0.26\\ 
  20.7 &   38.0 &   1.4 &   0.4 &  31.0 &   7.4 &  0.27\\ 
  20.7 &   47.5 &   1.5 &   0.5 &  19.6 &   8.4 &  0.25\\ 
  20.7 &   66.5 &   2.3 &   0.6 &  28.1 &   7.1 &  0.24\\ 
  20.7 &   76.0 &   2.3 &   0.7 &  28.8 &   8.3 &  0.23\\ 
  20.7 &   85.5 &   2.3 &   0.8 &  23.7 &   9.0 &  0.22\\ 
  20.7 &   95.0 &   2.7 &   1.2 &  -4.6 &  11.6 &  0.21\\ 
  20.7 &  104.5 &   2.9 &   1.2 & -12.8 &  10.9 &  0.18\\ 
  20.7 &  123.5 &   4.3 &   2.4 & -15.2 &  13.4 &  0.18\\ 
  10.8 &  -95.0 &   3.8 &   1.6 &  39.6 &  11.5 &  0.21\\ 
  10.8 &  -85.5 &   2.2 &   0.6 &  12.1 &   6.8 &  0.29\\ 
  10.8 &  -76.0 &   2.2 &   0.7 &  30.2 &   8.2 &  0.31\\ 
  10.8 &  -66.5 &   2.3 &   0.4 &  29.9 &   4.9 &  0.37\\ 
  10.8 &  -57.0 &   1.6 &   0.2 &  17.1 &   4.3 &  0.42\\ 
  10.8 &  -47.5 &   1.2 &   0.2 &   0.0 &   4.1 &  0.48\\ 
  10.8 &  -38.0 &   1.7 &   0.2 &  -5.0 &   3.5 &  0.43\\ 
  10.8 &  -28.5 &   2.1 &   0.3 &  -6.5 &   3.9 &  0.36\\ 
  10.8 &  -19.0 &   2.2 &   0.4 &   2.6 &   5.2 &  0.35\\ 
  10.8 &   -9.5 &   1.9 &   0.3 &   1.4 &   4.6 &  0.35\\ 
  10.8 &   -0.0 &   1.6 &   0.3 &   3.8 &   4.9 &  0.32\\ 
  10.8 &    9.5 &   1.2 &   0.3 &  14.8 &   7.4 &  0.30\\ 
  10.8 &   19.0 &   1.2 &   0.3 &  14.6 &   7.6 &  0.28\\ 
  10.8 &   28.5 &   1.5 &   0.4 &  18.6 &   6.2 &  0.27\\ 
  10.8 &   38.0 &   2.0 &   0.4 &  16.7 &   5.2 &  0.27\\ 
  10.8 &   47.5 &   1.8 &   0.4 &  22.6 &   6.0 &  0.27\\ 
  10.8 &   57.0 &   1.5 &   0.5 &  12.1 &   7.8 &  0.27\\ 
  10.8 &   66.5 &   2.1 &   0.6 &  14.3 &   7.3 &  0.27\\ 
  10.8 &   76.0 &   2.7 &   0.8 &  20.6 &   7.8 &  0.24\\ 
  10.8 &   85.5 &   1.6 &   0.6 &  23.6 &   9.9 &  0.25\\ 
  10.8 &   95.0 &   1.7 &   0.9 &   5.0 &  14.1 &  0.23\\ 
  10.8 &  114.0 &   3.4 &   1.4 &  27.3 &  10.4 &  0.20\\ 
   0.9 & -104.5 &   1.8 &   0.6 &  31.9 &   9.4 &  0.25\\ 
   0.9 &  -95.0 &   3.6 &   1.8 &  20.2 &  11.7 &  0.17\\ 
   0.9 &  -85.5 &   1.6 &   0.5 &  20.1 &   8.6 &  0.30\\ 
   0.9 &  -76.0 &   2.2 &   0.5 &   7.3 &   5.9 &  0.37\\ 
   0.9 &  -66.5 &   2.1 &   0.3 &   0.6 &   4.4 &  0.45\\ 
   0.9 &  -57.0 &   2.2 &   0.3 &  -5.0 &   3.9 &  0.47\\ 
   0.9 &  -47.5 &   2.1 &   0.2 &  -9.1 &   2.9 &  0.50\\ 
   0.9 &  -38.0 &   2.6 &   0.2 &  -6.7 &   2.0 &  0.44\\ 
   0.9 &  -28.5 &   2.4 &   0.2 &  -7.7 &   2.4 &  0.46\\ 
   0.9 &  -19.0 &   1.5 &   0.2 &  -2.0 &   4.8 &  0.45\\ 
   0.9 &   -9.5 &   1.6 &   0.2 &  -1.9 &   4.0 &  0.44\\ 
   0.9 &   -0.0 &   1.8 &   0.3 & -10.2 &   4.0 &  0.39\\ 
   0.9 &    9.5 &   1.1 &   0.3 &  -4.4 &   6.3 &  0.34\\ 
   0.9 &   19.0 &   1.1 &   0.3 &  -0.0 &   8.3 &  0.31\\ 
   0.9 &   38.0 &   1.4 &   0.3 &  10.5 &   6.6 &  0.29\\ 
   0.9 &   47.5 &   1.7 &   0.3 &   7.7 &   4.7 &  0.29\\ 
   0.9 &   57.0 &   1.5 &   0.3 &  18.0 &   5.3 &  0.29\\ 
   0.9 &   66.5 &   1.3 &   0.4 &  22.2 &   7.4 &  0.30\\ 
   0.9 &   76.0 &   0.8 &   0.4 &  16.2 &  13.2 &  0.33\\ 
   0.9 &   95.0 &   1.3 &   0.6 &  22.4 &  11.9 &  0.29\\ 
   0.9 &  114.0 &   5.1 &   2.0 &  49.0 &  10.1 &  0.21\\ 
   0.9 &  133.0 &  11.3 &   6.3 &  35.4 &  13.3 &  0.20\\ 
  -9.0 &  -76.0 &   2.9 &   0.7 &  -5.2 &   6.3 &  0.37\\ 
  -9.0 &  -66.5 &   1.9 &   0.5 &  -4.8 &   6.7 &  0.36\\ 
  -9.0 &  -57.0 &   1.9 &   0.4 & -11.2 &   6.2 &  0.36\\ 
  -9.0 &  -47.5 &   2.1 &   0.4 &  -5.9 &   5.5 &  0.39\\ 
  -9.0 &  -38.0 &   2.5 &   0.3 &  -8.1 &   3.1 &  0.46\\ 
  -9.0 &  -28.5 &   2.4 &   0.2 &  -9.0 &   2.0 &  0.57\\ 
  -9.0 &  -19.0 &   2.1 &   0.2 & -16.0 &   2.2 &  0.59\\ 
  -9.0 &   -9.5 &   1.8 &   0.2 & -10.3 &   2.8 &  0.53\\ 
  -9.0 &   -0.0 &   1.4 &   0.2 & -11.6 &   4.3 &  0.50\\ 
  -9.0 &    9.5 &   1.7 &   0.2 & -20.6 &   3.6 &  0.39\\ 
  -9.0 &   19.0 &   1.4 &   0.2 & -16.8 &   4.9 &  0.36\\ 
  -9.0 &   28.5 &   1.2 &   0.4 & -12.4 &   8.7 &  0.32\\ 
  -9.0 &   38.0 &   1.7 &   0.4 &  -6.0 &   6.5 &  0.32\\ 
  -9.0 &   57.0 &   1.2 &   0.3 &  15.0 &   7.3 &  0.33\\ 
  -9.0 &   66.5 &   1.0 &   0.3 &  11.7 &   9.0 &  0.36\\ 
  -9.0 &   76.0 &   1.1 &   0.2 & -26.6 &   5.1 &  0.50\\ 
  -9.0 &   85.5 &   1.1 &   0.2 & -16.8 &   4.4 &  0.72\\ 
  -9.0 &   95.0 &   1.3 &   0.3 &  16.7 &   7.3 &  0.46\\ 
  -9.0 &  114.0 &   2.8 &   1.4 &  30.0 &  12.6 &  0.23\\ 
  -9.0 &  133.0 &  14.3 &   7.6 &  38.0 &  11.7 &  0.20\\ 
 -18.9 & -123.5 &   6.9 &   3.1 & -60.0 &   9.9 &  0.28\\ 
 -18.9 &  -95.0 &   4.8 &   2.0 & -41.6 &   9.9 &  0.22\\ 
 -18.9 &  -76.0 &   5.2 &   2.5 & -39.7 &  11.4 &  0.22\\ 
 -18.9 &  -47.5 &   1.1 &   0.4 &   6.3 &   8.6 &  0.38\\ 
 -18.9 &  -38.0 &   1.7 &   0.2 &  -0.0 &   4.2 &  0.49\\ 
 -18.9 &  -28.5 &   1.8 &   0.2 &  -6.4 &   3.2 &  0.57\\ 
 -18.9 &  -19.0 &   2.3 &   0.2 & -17.2 &   2.0 &  0.61\\ 
 -18.9 &   -9.5 &   1.7 &   0.1 & -22.7 &   2.4 &  0.58\\ 
 -18.9 &   -0.0 &   1.4 &   0.2 & -27.4 &   3.3 &  0.57\\ 
 -18.9 &    9.5 &   1.3 &   0.2 & -26.6 &   3.4 &  0.56\\ 
 -18.9 &   19.0 &   1.5 &   0.2 & -28.3 &   3.1 &  0.47\\ 
 -18.9 &   28.5 &   1.3 &   0.2 & -20.1 &   4.8 &  0.41\\ 
 -18.9 &   38.0 &   1.2 &   0.2 & -16.4 &   5.4 &  0.38\\ 
 -18.9 &   47.5 &   0.9 &   0.2 & -15.7 &   6.6 &  0.41\\ 
 -18.9 &   57.0 &   1.0 &   0.2 &  -2.5 &   6.1 &  0.46\\ 
 -18.9 &   66.5 &   1.7 &   0.2 &  -2.4 &   4.1 &  0.53\\ 
 -18.9 &   76.0 &   1.1 &   0.2 &  -0.9 &   4.1 &  0.70\\ 
 -18.9 &   85.5 &   1.5 &   0.1 &   2.6 &   2.5 &  0.88\\ 
 -18.9 &   95.0 &   1.4 &   0.2 &  11.4 &   4.4 &  0.59\\ 
 -18.9 &  104.5 &   1.1 &   0.4 &  30.0 &   9.6 &  0.41\\ 
 -28.8 &  -85.5 &   3.2 &   1.2 & -77.2 &   9.3 &  0.28\\ 
 -28.8 &  -47.5 &   0.7 &   0.3 & -30.0 &   9.9 &  0.40\\ 
 -28.8 &  -38.0 &   1.1 &   0.3 & -17.1 &   7.0 &  0.45\\ 
 -28.8 &  -28.5 &   1.9 &   0.2 &  -7.3 &   3.8 &  0.44\\ 
 -28.8 &  -19.0 &   2.8 &   0.4 &  -7.8 &   3.7 &  0.44\\ 
 -28.8 &   -9.5 &   1.4 &   0.2 & -19.2 &   3.9 &  0.54\\ 
 -28.8 &   -0.0 &   1.1 &   0.2 & -21.3 &   3.9 &  0.58\\ 
 -28.8 &    9.5 &   1.1 &   0.2 & -28.9 &   4.6 &  0.60\\ 
 -28.8 &   19.0 &   1.1 &   0.2 & -34.4 &   4.6 &  0.58\\ 
 -28.8 &   28.5 &   0.9 &   0.2 & -34.5 &   5.6 &  0.56\\ 
 -28.8 &   38.0 &   1.1 &   0.3 & -16.1 &   6.7 &  0.44\\ 
 -28.8 &   47.5 &   0.7 &   0.2 &  -9.6 &   6.5 &  0.54\\ 
 -28.8 &   57.0 &   0.9 &   0.1 &  11.8 &   4.3 &  0.65\\ 
 -28.8 &   66.5 &   1.0 &   0.1 &  -0.6 &   3.1 &  0.90\\ 
 -28.8 &   76.0 &   1.0 &   0.1 &  -9.7 &   3.7 &  0.93\\ 
 -28.8 &   85.5 &   1.4 &   0.1 &   3.9 &   3.2 &  0.89\\ 
 -28.8 &   95.0 &   1.1 &   0.2 &  13.5 &   4.1 &  0.78\\ 
 -28.8 &  104.5 &   1.1 &   0.3 &  33.0 &   7.9 &  0.57\\ 
 -38.7 & -104.5 &   1.7 &   0.6 & -32.6 &   9.9 &  0.64\\ 
 -38.7 &  -76.0 &   4.1 &   1.5 & -70.3 &   9.3 &  0.25\\ 
 -38.7 &  -66.5 &   1.6 &   1.0 & -81.5 &  13.9 &  0.27\\ 
 -38.7 &  -38.0 &   1.0 &   0.3 & -20.1 &   9.5 &  0.35\\ 
 -38.7 &  -28.5 &   1.6 &   0.3 & -12.9 &   5.3 &  0.35\\ 
 -38.7 &  -19.0 &   2.7 &   0.3 & -10.7 &   2.8 &  0.36\\ 
 -38.7 &   -9.5 &   2.2 &   0.2 &  -8.2 &   2.7 &  0.45\\ 
 -38.7 &   -0.0 &   1.4 &   0.1 & -16.0 &   2.7 &  0.60\\ 
 -38.7 &    9.5 &   1.2 &   0.1 & -33.0 &   3.1 &  0.63\\ 
 -38.7 &   19.0 &   1.4 &   0.2 & -33.0 &   3.3 &  0.61\\ 
 -38.7 &   28.5 &   0.8 &   0.1 & -40.1 &   4.8 &  0.75\\ 
 -38.7 &   38.0 &   0.9 &   0.2 & -44.4 &   5.6 &  0.68\\ 
 -38.7 &   47.5 &   0.6 &   0.2 & -25.3 &  11.5 &  0.59\\ 
 -38.7 &   76.0 &   0.9 &   0.2 &  17.2 &   6.2 &  0.79\\ 
 -38.7 &   85.5 &   0.8 &   0.3 &  18.0 &  10.1 &  0.73\\ 
 -38.7 &   95.0 &   1.2 &   0.2 &  16.6 &   3.6 &  0.88\\ 
 -38.7 &  104.5 &   0.6 &   0.2 &  19.4 &   8.9 &  0.79\\ 
 -38.7 &  123.5 &   4.7 &   2.3 &  64.5 &  11.9 &  0.27\\ 
 -48.6 & -104.5 &   2.3 &   0.9 & -60.6 &   9.8 &  0.49\\ 
 -48.6 &  -95.0 &   4.7 &   0.9 & -35.1 &   4.9 &  0.39\\ 
 -48.6 &  -76.0 &   4.4 &   1.8 & -75.2 &  10.3 &  0.21\\ 
 -48.6 &  -66.5 &   2.7 &   1.2 & -70.9 &  10.9 &  0.26\\ 
 -48.6 &  -57.0 &   3.5 &   0.9 & -72.1 &   7.4 &  0.26\\ 
 -48.6 &  -38.0 &   1.0 &   0.3 & -34.5 &   9.3 &  0.34\\ 
 -48.6 &  -28.5 &   1.4 &   0.3 & -22.5 &   7.2 &  0.35\\ 
 -48.6 &  -19.0 &   2.6 &   0.4 &  -9.4 &   3.8 &  0.36\\ 
 -48.6 &   -9.5 &   1.9 &   0.2 &  -9.1 &   3.6 &  0.41\\ 
 -48.6 &   -0.0 &   1.7 &   0.1 & -17.5 &   2.1 &  0.59\\ 
 -48.6 &    9.5 &   1.3 &   0.1 & -26.2 &   2.1 &  0.68\\ 
 -48.6 &   19.0 &   0.9 &   0.1 & -22.8 &   3.7 &  0.66\\ 
 -48.6 &   28.5 &   0.5 &   0.1 & -34.9 &   5.0 &  0.71\\ 
 -48.6 &   38.0 &   0.7 &   0.1 & -43.1 &   4.3 &  0.68\\ 
 -48.6 &   47.5 &   0.5 &   0.1 & -38.9 &   6.7 &  0.60\\ 
 -48.6 &   57.0 &   0.3 &   0.1 & -29.3 &  12.9 &  0.63\\ 
 -48.6 &   66.5 &   1.0 &   0.2 &   4.0 &   4.4 &  0.73\\ 
 -48.6 &   76.0 &   0.7 &   0.2 & -19.5 &   8.1 &  0.63\\ 
 -48.6 &   95.0 &   1.3 &   0.3 &  26.2 &   6.1 &  0.71\\ 
 -48.6 &  104.5 &   1.2 &   0.3 &  28.0 &   7.8 &  0.60\\ 
 -58.5 & -114.0 &   5.3 &   1.7 & -60.7 &   8.0 &  0.33\\ 
 -58.5 & -104.5 &   2.9 &   0.6 & -52.5 &   5.3 &  0.61\\ 
 -58.5 &  -95.0 &   3.5 &   0.6 & -38.3 &   4.6 &  0.53\\ 
 -58.5 &  -85.5 &   5.3 &   1.2 & -39.5 &   5.8 &  0.35\\ 
 -58.5 &  -76.0 &   5.6 &   2.3 & -57.0 &   9.7 &  0.24\\ 
 -58.5 &  -57.0 &   3.0 &   0.9 & -72.4 &   8.0 &  0.26\\ 
 -58.5 &  -47.5 &   1.7 &   0.8 & -65.5 &  11.6 &  0.27\\ 
 -58.5 &  -38.0 &   1.0 &   0.5 & -25.1 &  13.3 &  0.29\\ 
 -58.5 &  -28.5 &   2.0 &   0.4 & -36.4 &   5.3 &  0.32\\ 
 -58.5 &  -19.0 &   2.5 &   0.4 & -16.8 &   4.0 &  0.35\\ 
 -58.5 &   -9.5 &   2.6 &   0.3 &  -9.0 &   3.4 &  0.39\\ 
 -58.5 &   -0.0 &   1.5 &   0.2 & -11.3 &   3.7 &  0.53\\ 
 -58.5 &    9.5 &   1.0 &   0.1 &  -7.4 &   3.0 &  0.83\\ 
 -58.5 &   19.0 &   0.4 &   0.1 & -16.8 &   7.6 &  0.82\\ 
 -58.5 &   28.5 &   0.5 &   0.1 & -30.6 &   6.3 &  0.86\\ 
 -58.5 &   38.0 &   0.7 &   0.1 & -46.2 &   5.5 &  0.75\\ 
 -58.5 &   47.5 &   0.7 &   0.2 & -58.6 &   6.9 &  0.60\\ 
 -58.5 &   57.0 &   1.3 &   0.2 & -30.0 &   4.6 &  0.49\\ 
 -58.5 &   66.5 &   2.5 &   0.3 &  -7.3 &   3.3 &  0.44\\ 
 -58.5 &   76.0 &   1.2 &   0.3 & -11.5 &   7.6 &  0.46\\ 
 -68.4 & -114.0 &   8.9 &   2.0 & -49.6 &   4.6 &  0.49\\ 
 -68.4 & -104.5 &   5.3 &   0.8 & -46.2 &   3.5 &  0.63\\ 
 -68.4 &  -95.0 &   3.4 &   0.5 & -43.5 &   3.9 &  0.65\\ 
 -68.4 &  -85.5 &   3.8 &   0.7 & -53.4 &   4.8 &  0.47\\ 
 -68.4 &  -76.0 &   3.7 &   1.1 & -45.5 &   7.5 &  0.40\\ 
 -68.4 &  -66.5 &   3.3 &   1.4 & -59.7 &  10.0 &  0.26\\ 
 -68.4 &  -57.0 &   2.0 &   1.1 & -63.7 &  13.2 &  0.26\\ 
 -68.4 &  -47.5 &   2.4 &   1.0 & -51.0 &  10.2 &  0.26\\ 
 -68.4 &  -38.0 &   1.9 &   0.7 & -60.8 &   9.2 &  0.28\\ 
 -68.4 &  -28.5 &   2.1 &   0.4 & -37.4 &   5.6 &  0.33\\ 
 -68.4 &  -19.0 &   1.9 &   0.3 & -24.9 &   4.0 &  0.39\\ 
 -68.4 &   -9.5 &   1.8 &   0.2 & -22.1 &   3.8 &  0.41\\ 
 -68.4 &   -0.0 &   1.5 &   0.2 & -19.6 &   2.9 &  0.55\\ 
 -68.4 &    9.5 &   0.9 &   0.1 & -18.6 &   3.0 &  0.79\\ 
 -68.4 &   19.0 &   0.8 &   0.1 & -14.5 &   3.2 &  0.83\\ 
 -68.4 &   28.5 &   0.5 &   0.1 &  -2.3 &   5.3 &  0.85\\ 
 -68.4 &   38.0 &   0.3 &   0.1 & -65.3 &  11.3 &  0.78\\ 
 -68.4 &   47.5 &   0.9 &   0.2 & -73.6 &   5.2 &  0.54\\ 
 -68.4 &   57.0 &   1.1 &   0.2 & -58.7 &   6.3 &  0.43\\ 
 -68.4 &   66.5 &   1.7 &   0.4 & -31.7 &   6.1 &  0.36\\ 
 -68.4 &   76.0 &   1.8 &   0.5 & -16.5 &   7.3 &  0.34\\ 
 -68.4 &   85.5 &   3.1 &   1.1 & -22.1 &   8.8 &  0.29\\ 
 -78.3 & -104.5 &   5.8 &   1.9 & -45.4 &   7.6 &  0.68\\ 
 -78.3 &  -95.0 &   6.4 &   1.0 & -42.8 &   3.7 &  0.58\\ 
 -78.3 &  -85.5 &   4.6 &   0.8 & -44.0 &   4.0 &  0.53\\ 
 -78.3 &  -76.0 &   5.0 &   1.1 & -52.9 &   5.2 &  0.36\\ 
 -78.3 &  -47.5 &   2.1 &   1.2 & -53.1 &  13.7 &  0.25\\ 
 -78.3 &  -38.0 &   4.4 &   0.8 & -67.3 &   4.9 &  0.28\\ 
 -78.3 &  -28.5 &   2.4 &   0.7 & -62.5 &   8.3 &  0.31\\ 
 -78.3 &  -19.0 &   1.2 &   0.5 & -40.6 &   9.5 &  0.43\\ 
 -78.3 &   -9.5 &   1.3 &   0.4 & -41.3 &   7.6 &  0.44\\ 
 -78.3 &   -0.0 &   0.6 &   0.3 & -24.8 &  11.3 &  0.51\\ 
 -78.3 &    9.5 &   1.4 &   0.2 & -16.2 &   4.4 &  0.58\\ 
 -78.3 &   19.0 &   0.5 &   0.2 &  15.6 &  10.9 &  0.66\\ 
 -78.3 &   28.5 &   0.8 &   0.2 &  -1.4 &   5.3 &  0.70\\ 
 -78.3 &   66.5 &   1.9 &   0.5 & -35.2 &   6.8 &  0.33\\ 
 -78.3 &   76.0 &   1.7 &   0.6 & -52.9 &   9.5 &  0.31\\ 
 -88.2 &  -95.0 &  10.0 &   2.6 & -49.6 &   4.8 &  0.36\\ 
 -88.2 &  -85.5 &   5.4 &   1.6 & -52.6 &   6.7 &  0.53\\ 
 -88.2 &  -76.0 &   5.8 &   1.9 & -50.9 &   6.0 &  0.42\\ 
 -88.2 &  -66.5 &   4.9 &   1.8 & -42.9 &   8.5 &  0.28\\ 
 -88.2 &  -57.0 &   3.0 &   1.6 & -50.3 &  12.8 &  0.28\\ 
 -88.2 &  -47.5 &   5.6 &   2.0 & -54.5 &   7.8 &  0.25\\ 
 -88.2 &  -38.0 &   3.5 &   0.9 & -56.0 &   6.8 &  0.27\\ 
 -88.2 &  -28.5 &   2.6 &   0.8 & -38.1 &   8.3 &  0.28\\ 
 -88.2 &  -19.0 &   2.3 &   0.8 & -22.1 &   9.5 &  0.31\\ 
 -88.2 &   -9.5 &   3.1 &   0.7 & -25.9 &   6.3 &  0.32\\ 
 -88.2 &   -0.0 &   2.4 &   0.6 & -43.5 &   7.9 &  0.35\\ 
 -88.2 &   28.5 &   0.5 &   0.3 & -44.4 &  13.7 &  0.56\\ 
 -88.2 &   47.5 &   1.5 &   0.4 & -53.7 &   7.8 &  0.44\\ 
 -88.2 &   57.0 &   3.1 &   1.4 & -88.5 &  11.1 &  0.42\\ 
 -88.2 &   66.5 &   1.8 &   0.9 & -57.7 &  13.3 &  0.32\\ 
 -88.2 &   76.0 &   2.3 &   1.2 &  89.6 &  12.4 &  0.37\\ 
 -88.2 &   85.5 &   2.0 &   1.0 & -71.7 &  12.0 &  0.39\\ 
 -98.1 &  -66.5 &   3.3 &   1.7 & -48.9 &  12.1 &  0.29\\ 
 -98.1 &  -57.0 &   4.0 &   1.4 & -51.4 &   8.9 &  0.29\\ 
 -98.1 &  -38.0 &   3.8 &   1.5 & -63.4 &  10.4 &  0.27\\ 
 -98.1 &  -19.0 &   1.8 &   1.0 & -18.9 &  13.4 &  0.30\\ 
 -98.1 &   -9.5 &   3.4 &   1.0 & -20.3 &   7.8 &  0.32\\ 
 -98.1 &   -0.0 &   2.3 &   0.6 & -44.2 &   7.5 &  0.35\\ 
 -98.1 &    9.5 &   1.0 &   0.6 & -46.5 &  15.2 &  0.36\\ 
 -98.1 &   19.0 &   1.3 &   0.7 & -12.7 &  12.0 &  0.40\\ 
 -98.1 &   47.5 &   1.2 &   0.3 & -46.7 &   8.6 &  0.44\\ 
 -98.1 &   57.0 &   1.4 &   0.5 & -36.6 &  11.4 &  0.43\\ 
 -98.1 &   76.0 &   3.6 &   2.0 & -80.0 &  10.1 &  0.43\\ 
 -98.1 &   85.5 &   3.8 &   2.1 & -66.1 &  11.1 &  0.47\\
\enddata
\tablenotetext{a}{Offsets are given with respect to $18^\text{h} 20^\text{m} 25^\text{s}$, $-16\arcdeg 13\arcmin 02\arcsec$ (J2000).} 
\tablenotetext{b}{Position angle of electric vector measured east from north.}
\tablenotetext{c}{Intensity normalized to 1.00 at peak.}
%\label{table:pvectors}
\end{deluxetable}

\end{document}